\documentclass[11pt,a4paper]{article}
\pdfoutput=1
\usepackage{jheppub} 
\usepackage{xspace}
\usepackage{subfigure}
\usepackage{amsmath}
\usepackage{amssymb}
\usepackage{amsfonts}
\usepackage{mathrsfs}

\newcommand{\be}{\begin{equation}}
\newcommand{\ee}{\end{equation}}

\newcommand{\bea}{\begin{eqnarray}}
\newcommand{\eea}{\end{eqnarray}}
\newcommand{\beanon}{\begin{eqnarray*}}
\newcommand{\eeanon}{\end{eqnarray*}}
\newcommand{\ba}{\begin{array}}
\newcommand{\ea}{\end{array}}
\newcommand{\bd}{\begin{description}}
\newcommand{\ed}{\end{description}}
\newcommand{\bi}{\begin{itemize}}
\newcommand{\ei}{\end{itemize}}
\newcommand{\ben}{\begin{enumerate}}
\newcommand{\een}{\end{enumerate}}
\newcommand{\bc}{\begin{center}}
\newcommand{\ec}{\end{center}}

\newcommand{\ordEW}{\mathcal{O}(\alpha_{\scriptscriptstyle EM}^6)\xspace}
\newcommand{\ordQCD}{\mathcal{O}(\alpha_{\scriptscriptstyle EM}^4
  \alpha_{\scriptscriptstyle S}^2)\xspace}

\newcommand{\eqn}[1]{eq.(\ref{#1})}

\newcommand{\tbn}[1]{tab.~\ref{#1}}

\newcommand{\fig}[1]{fig.~\ref{#1}}

\newcommand{\rf}[1]{ref.~\cite{#1}}
\newcommand{\rfs}[1]{refs.~\cite{#1}}  % comma separated list

\newcommand{\Phantom}{{\tt PHANTOM}\xspace}

\newcommand{\sla}[1]{\displaystyle{\not} #1}

\newcommand{\CP}{\ensuremath{\mathcal{C}\mathcal{P}}\xspace}

\newcommand{\al}{\alpha}

\newcommand{\hzero}{\ensuremath{h}} %light neutral Higgs
\newcommand{\Hzero}{\ensuremath{H}} %light neutral Higgs

\title{
Interference Effects in Higgs production through Vector Boson Fusion in the Standard Model and its
Singlet Extension.
}

\author[a]{Alessandro Ballestrero}
\author[a,b]{and Ezio Maina}

\affiliation[a]{INFN, Sezione di Torino,\\
Via Giuria 1, 10125 Torino, Italy}
\affiliation[b]{Dipartimento di Fisica, Universit\`a di Torino,\\
Via Giuria 1, 10125 Torino, Italy}

\emailAdd{ballestrero@to.infn.it}
\emailAdd{maina@to.infn.it}

\abstract{
Interference effects play an important role in Electroweak Physics.
They are responsible for the restoration of unitarity al large energies.
When, as is often the case, higher order corrections are only available
for some particular subamplitude, interferences need to be carefully
computed in order to obtain the best theoretical prediction.
In the new proposal to estimate the total Higgs width from the off shell
cross section, the interference between the Higgs signal and the
background is essential.
It has been recently pointed out in gluon fusion that whenever more than
one neutral, \CP even, scalars are present in the spectrum large cancellations
can occur.
We extend these studies to Vector Boson Scattering, examining interference
effects in the Higgs sector in the Standard Model and its one Higgs Singlet extension.
}

\begin{document}

\maketitle

% There is a problem with five digits arXiv numbers in JHEP.bst
        
\section{Introduction}
\label{sec:intro}
Now that a resonance has been dicovered at about 125 GeV  \cite{Aad:2012tfa,Chatrchyan:2012ufa},
the race is on to measure all its properties.
All studies  based on LHC Run I data are consistent with the hypothesis that
the new particle is indeed the Standard Model Higgs boson.
The mass is already known with an uncertainty of  two per mill from the latest
published analyses \cite{Khachatryan:2014jba,Aad:2015zhl} and the signal strengths
$\mu^i = \sigma^i/\sigma^i_{SM}$, where $i$ runs over the decay channnels, 
are known to about 10 to 20\% \cite{Khachatryan:2014jba,ATLAS:2014yyy,ATLAS:2014bny}.
There is still room for more complicated Higgs sectors but compatibility 
with experimental results is severely restricting their parameter space \cite{CMS:2014www}. 
In Run II,  larger luminosity and energy will provide more precise measurements of the
characteristics of the new particle and extend the mass range in which other scalars can be searched for.

Lately, a lot of attention has been paid to the prospects of detailed studies of off-shell Higgs
contributions. On the one hand, at large energies, Higgs exchange unitarizes processes like
Vector Boson Scattering (VBS) and fermion pair annihilation to Vector Bosons which would otherwise
diverge. On the other hand, the comparison of off-shell and peak cross sections can provide limits on the
total width of the Higgs \cite{Caola:2013yja},
 exploiting the interference of the Higgs contribution with the rest of the
amplitude and the different dependence on the Higgs couplings of the two terms.
Both aspects are sensitive to BSM physics, both through direct production of new states and through
their contributions in loops.
 
If additional neutral scalars are present in the physical spectrum, non trivial interference effects have been
demonstrated in Gluon Gluon Fusion (GGF) processes
\cite{Englert:2014ffa,Maina:2015ela,Kauer:2015hia,Englert:2015zra}. 

It is quite natural to extend these studies to VBS which has been traditionally regarded as the ultimate
testing ground of the ElectroWeak Symmetry Breaking mechanism. The ratio of the Higgs production 
cross section in Vector Boson Fusion (VBF) to the cross section in gluon fusion grows for larger
Higgs masses and, as a consequence, the importance of VBF as a discovery channel for new scalar
resonances of an extended Higgs sector increases.
VBF is not affected by BSM physics through loops\cite{Englert:2014aca}, therefore it can be argued that
the limits it provides on the Higgs width are less model dependent than those obtained in GGF.
It is well known that interference effects between Higgs exchange diagrams and all other ones are large in
VBF.
The interference between Higgs fields of different masses in VBS will also be present and
modulate the cancellations which restore unitarity.

There is a
widespread belief that accurate predictions for the production af a heavy Higgs can be
obtained by computing $pp \rightarrow jjH$, possibly folded with a Breit-Wigner distribution in
order to control the effects of the Higgs width, and then decaying $H$ to the desired final state.
The appeal of this point of view is that higher order ElectroWeak  corrections to  
$pp \rightarrow jjH$ are available at NLO \cite{Ciccolini:2007jr,Ciccolini:2007ec} and QCD
corrections are known almost exactly at NNLO  \cite{Bolzoni:2010xr,Bolzoni:2011cu}.
However there are large interference effects among Higgs exchange diagrams
already in the SM and similar phenomena  are to be expected
between the SM like Higgs and its eventual heavier partner, producing non
negligible modifications to the cross section and resonance shape of the latter.

Run II will certainly allow to study Vector Boson Fusion in greater detail than it was possible with the
limited statistics collected in Run I.

Since the landscape of possible extensions of the SM Higgs sector is quite complicated, it makes sense to
examine the simplest renormalizable enlargement, that is the one Higgs Singlet Model (1HSM).
It introduces one 
additional real scalar field which is a singlet under all SM gauge groups. 
The 1HSM has been extensively investigated in the literature 
\cite{Silveira:1985rk,Schabinger:2005ei,O'Connell:2006wi,BahatTreidel:2006kx,
Barger:2007im,Bhattacharyya:2007pb,Gonderinger:2009jp,Dawson:2009yx,Bock:2010nz,
Fox:2011qc,Englert:2011yb,Englert:2011us,Batell:2011pz,Englert:2011aa,Gupta:2011gd,
Batell:2012mj,No:2013wsa,Pruna:2013bma,Lopez-Val:2014jva,Englert:2014ffa,Profumo:2014opa,
Chen:2014ask,Robens:2015gla,
Logan:2014ppa,Maina:2015ela,Falkowski:2015iwa,Kauer:2015hia,Englert:2015zra}.
Recently, a great deal of activity has concentrated on establishing the restrictions imposed on its
parameter space by theoretical and experimental constraints \cite{Pruna:2013bma,Lopez-Val:2014jva,
Robens:2015gla,Falkowski:2015iwa}; on interference effects between the two neutral Higgs fields and
with the continuum
\cite{Maina:2015ela,Kauer:2015hia,Englert:2015zra} and on possible consequences on the determination
of the Higgs width through a measurement of the off-shell Higgs cross section \cite{Englert:2014ffa,
Logan:2014ppa}, as proposed in \rf{Caola:2013yja}.

To the best of our knowledge, all analyses
so far have resorted to a superposition of a VBF Higgs signal times decay sample to the continuum,
ignoring interferences, since no public MC is available for VBS in the 1HSM. 
We have upgraded \Phantom \cite{Ballestrero:2007xq},
allowing for the simulation of the 1HSM and more generally for the presence of two neutral scalars.

In this paper we apply this new tool to study interference effects in 
$p p \rightarrow jj l^+l^-l^{\prime +}l^{\prime -}$ and
$p p \rightarrow jj l^+\bar{\nu}_l l^{\prime -}\nu_{l^\prime}$ production,
where both $l$ and $l^{\prime}$ can be either an electron or a muon,  $l\neq l^{\prime}$.
This is a case study rather then a complete analysis and we are aware that
rates are expected to be small \cite{Ballestrero:2010vp,Campbell:2015vwa}. A careful investigation of all
channels, including 
the semileptonic ones and exploiting all techniques to identify vector bosons decaying hadronically will be
required to assess the observability of the 1HSM through VBF in Run II and beyond.

\section{The Singlet Extension of the Standard Model}
\label{sec:model}
In the following we consider the singlet extension of the SM in the notation
of \rf{Pruna:2013bma}. A real {$SU(2)_L\otimes\,U(1)_Y$} singlet, $S$, is introduced and the term:
\begin{equation}\label{lag:s}
\mathscr{L}_s = 
\partial^{\mu} S \partial_{\mu} S 
 -\mu_1^2 \Phi^{\dagger} \Phi -\mu_2^2 S ^2 + \lambda_1
(\Phi^{\dagger} \Phi)^2 + \lambda_2  S^4 + \lambda_3 \Phi^{\dagger}
\Phi S ^2.
\end{equation}
is added to the SM Lagrangian, where $\Phi$ is the usual Higgs doublet. 
$\mathscr{L}_s $ is gauge invariant and renormalizable.
A $\mathcal{Z}_2$ symmetry , $S \leftrightarrow -S$,
which forbids additional terms in the potential is assumed. A detailed discussion of the 1HSM without
$\mathcal{Z}_2$ symmetry can be found in \rfs{O'Connell:2006wi,Barger:2007im,No:2013wsa,
Profumo:2014opa,Chen:2014ask}.

The neutral components of these fields can be expanded around their
respective Vacuum Expectation Values:

\begin{equation}
 \Phi = \left(\begin{array}{c}  
  G^{\pm} \\ \cfrac{v_d + l^0 + iG^0}{\sqrt{2}}
 \end{array}\right) \qquad \qquad S = \cfrac{v_s + s^0}{\sqrt{2}}
 \label{eq:components}.
\end{equation}

\noindent The minimum of the potential is achieved for 

\begin{alignat}{5}
 \mu^2_1 = \lambda_1 v_d^2 + \cfrac{\lambda_3 v_s^2}{2}; \qquad \qquad 
  \mu^2_2 = \lambda_2 v_s^2 + \cfrac{\lambda_3 v^2_d}{2}
  \label{eq:minimum},
\end{alignat}
provided
\begin{alignat}{5}
 \lambda_1, \lambda_2 > 0; \qquad 4\lambda_1\lambda_2 - \lambda_3^2 > 0 \label{eq:stability}\; .
\end{alignat}

The  mass matrix can be diagonalized introducing new fields $\hzero$ and $\Hzero$:
\begin{alignat}{5}
\hzero =l^0  \cos\alpha  -s^0 \sin\alpha  \qquad \text{and} \qquad
 \Hzero &= l^0 \sin\alpha  +s^0 \cos\alpha 
\label{eq:masseigen-repeat}
\end{alignat}
with $-\frac{\pi}{2} < \alpha < \frac{\pi}{2}$.

\noindent The  masses are

\begin{equation}
 M^2_{\hzero,\Hzero} = \lambda_1\,v_d^2 + 
 \lambda_2\,v_s^2 \mp |\lambda_1\,v_d^2 - 
 \lambda_2\,v_s^2|\,\sqrt{1+\tan^2(2\alpha)}\, ,
\quad 
 \tan(2\alpha) = \cfrac{\lambda_3v_dv_s}{\lambda_1 v_d^2 - \lambda_2v_s^2}\, ,
 \label{eq:masseigen}
\end{equation}
\noindent  with the convention $M_{\Hzero}^2 > M_{\hzero}^2$.

The Higgs sector in this model is determined by five independent parameters,  which can be chosen as
\begin{equation}
\label{eq:parameters}
m_{\hzero},\,m_{\Hzero}, \,\sin\al, \,{v_d}, \,\tan\beta \,\equiv\,{v_d}/{v_s}\, ,
\end{equation} 
where the doublet VEV is fixed in terms of the Fermi constant through $v_d^2 = G_F^{-1}/\sqrt{2}$. 
Furthermore one of the Higgs masses is determined by the LHC measurement of $125.02$ GeV.
Therefore, three parameters of the model, $M_{\Hzero}, \,\sin\al, \,\tan\beta $, are at present 
undetermined. 

The Feynman rules for the 1HSM have been derived using FeynRules 
\cite{Christensen:2008py,Alloul:2013bka}.  
\footnote{The corresponding UFO file \cite{Degrande:2011ua}, 
which allows the simulation at tree level of any process in the model,
can be downloaded from http://personalpages.to.infn.it/$\scriptstyle \sim$maina/Singlet.
}

It should be mentioned that allowing a discrete symmetry to be spontaneously broken, as is the case in
the simplified model considered here when the singlet field $S$ has a non zero vacuum expectation
value,  will introduce potentially problematic cosmic domain walls
\cite{Zeldovich:1974uw,Kibble:1976sj,Kibble:1980mv,Abel:1995wk,Friedland:2002qs,Barger:2008jx}.
These considerations, however have little bearing on the paper's main point.

For future reference, we report the expression of the tree level partial width for the decay of the heavy 
scalar into two light ones:
\begin{equation}
\label{eq:H2hhWidth}
\Gamma (H\rightarrow h h ) = \frac{e^2 M_H^3}{128 \pi M_W^2 s_W^2}
\left( 1 - \frac{4M_h^2}{M_H^2} \right)^{\frac{1}{2}}
\left( 1 + \frac{2M_h^2}{M_H^2} \right)^2
s_{\alpha}^2 c_{\alpha}^2 \left( c_{\alpha} + s_{\alpha} \tan\beta \right) ^2
\end{equation}

and those of the width of both scalars: 
\begin{equation}
\label{eq:scalarWidths}
\Gamma_h = \Gamma^{SM}(M_h) c_\alpha^2, \qquad 
\Gamma_H = \Gamma^{SM}(M_H) s_\alpha^2 + \Gamma(H\rightarrow hh)
\end{equation}
where $c_\alpha = \cos\alpha,\,s_\alpha = \sin\alpha$.

%\section{Limits on the parameters}
%\label{sec:limits}
The strongest limits on the parameters of the 1HSM 
\rf{Lopez-Val:2014jva,Robens:2015gla,Falkowski:2015iwa}
come from measurements of the coupling strengths of the light Higgs 
\cite{Khachatryan:2014jba,ATLAS:2014yyy,ATLAS:2014bny,CMS:2014www},
which dominate for small masses of the heavy Higgs,
and from the contribution of higher order corrections to precision measurements, in
particular to the mass of the W boson \cite{Lopez-Val:2014jva}, which provides the tightest constraint
for large $M_H$.
The most precise result for the overall coupling strength of the Higgs boson from CMS
\cite{Khachatryan:2014jba} reads

\begin{equation}
\hat{\mu} = \hat{\sigma}/\sigma_{SM} = 1.00 \pm 0.13.
\end{equation}
Therefore the absolute value of $\sin\alpha$ cannot be larger than about 0.4. This is in agreement with
the limits obtained in \rf{Lopez-Val:2014jva,Robens:2015gla,Falkowski:2015iwa} 
which conclude that the largest possible
value for the absolute value of $\sin\alpha$ is 0.46 for $M_H$ between 160 and 180 GeV.  This limit
becomes slowly more stringent for increasing heavy Higgs masses reaching about 0.2 at $M_H = 700$
GeV.

\section{New Features in \Phantom}
\label{sec:phantom}

\Phantom  has been upgraded to allow for the presence of two neutral 
\CP even scalars.
The parameters which control how the Higgs sector is simulated, with masses and widths expressed in
GeV, are:
\begin{itemize}
\item \texttt{rmh}: light Higgs mass. If \texttt{rmh} $<$  0 all light and heavy Higgs exchange diagrams
are set to zero.
\item \texttt{gamh}: light Higgs width.  If \texttt{gamh} $<$ 0 the width is computed internally
following the prescription of \rf{Heinemeyer:2013tqa} and multiplied by $\cos^2\alpha$ if working
in the 1HSM. 
 \end{itemize}
The parameter \texttt{i\_singlet} selects whether \Phantom performs the calculations in the 
 SM (\texttt{i\_singlet}=0) or in the 1HSM (\texttt{i\_singlet}=1). 
If the 1HSM is selected the following inputs are required:
\begin{itemize}
\item \texttt{rmhh}: heavy Higgs mass.  If \texttt{rmhh} $<$  0 all heavy Higgs exchange diagrams
are set to zero.
\item \texttt{rcosa}: the cosine of the mixing angle $\alpha$.
\item \texttt{tgbeta}: $\tan\beta$.
\item \texttt{gamhh}: heavy Higgs width. If \texttt{gamhh} $<$ 0 the width is computed internally
following the prescription of \rf{Heinemeyer:2013tqa}  and then multiplied by $\sin^2\alpha$. 
$\Gamma (H\rightarrow h h )$, \eqn{eq:H2hhWidth}, is then added to the result.
\end{itemize}
Moreover the contribution of the Higgs exchange diagrams
can be computed separately, both in the SM and in the 1HSM, setting the following flag:
\begin{itemize}
\item \texttt{i\_signal}: if \texttt{i\_signal} = 0 the full matrix element is computed.\\
If \texttt{i\_signal} $>$ 0 only a set of Higgs exchange diagrams are evaluated at $\ordEW$:
      \begin{itemize}
				\item   \texttt{i\_signal} = 1: $s$-channel exchange contributions. 
				\item   \texttt{i\_signal} = 2:  all Higgs exchange contributions to VV scattering. 
				\item   \texttt{i\_signal} = 3:  all Higgs exchange contributions to VV scattering plus the
				Higgsstrahlung diagrams with $h,H \rightarrow VV$.
     \end{itemize}
\end{itemize}

\section{Notation and details of the calculation}
\label{sec:notation}

We are going to present results, at the 13 TeV LHC,  for
$p p \rightarrow jj\, l^+l^-l^{\prime +}l^{\prime -}$
and $p p \rightarrow jj\, l^+\bar{\nu}_l l^{\prime -}\nu_{l^\prime}$  production,
where $l(l^{\prime}) = e,\mu$, $l\neq l^{\prime}$. We have identified the light Higgs $h$ with the
resonance discovered  in Run I and set its mass to 125 GeV, concentrating on the scenario in which
the heavy Higgs $H$ is still undetected.

Samples of events have been generated with \Phantom
using CTEQ6L1 parton distribution functions \cite{Pumplin:2002vw}.
The ratio of vacuum expectation values, $\tan\beta$, has been taken equal to 0.3 for
$M_H = 600$ GeV and $M_H = 900$ GeV,  and equal to 1.0 for $M_H = 400$ GeV.
This corresponds, using \eqn{eq:H2hhWidth} for the $H\rightarrow hh$ width and 
\rf{Heinemeyer:2013tqa} for the SM Higgs width,
to $\Gamma_H = 4.08$ GeV
for $M_H = 400$ GeV, $s_{\alpha} = 0.3$; $\Gamma_H = 6.45$ GeV
for $M_H = 600$ GeV and $s_{\alpha} = 0.2$;
$\Gamma_H = 89.14$ GeV for $M_H = 900$ GeV and $s_{\alpha} = 0.4$.

The charged leptons are required to satisfy:
\begin{equation}
\label{eq:cuts_lep}
p_{Tl} >   20\: \mbox{GeV} ,\qquad  \vert\eta_l\vert <     3.0  ,\qquad  m_{l^+ l^-}  > 20 \: \mbox{GeV} 
\end{equation}
while the cuts on the jets are:    
\begin{equation}
\label{eq:cuts_jet}
p_{Tj} >   20 \: \mbox{GeV} ,\qquad  \vert\eta_j\vert <     6.5  ,
\qquad  m_{j_1j_2}   > 400 \: \mbox{GeV},
\qquad  \Delta\eta_{j_1j_2} > 2.0.
\end{equation}

For proceses with two charged leptons and two neutrinos in the final state we further impose:
\begin{equation}
\sla{p_{T}} >   20 \: \mbox{GeV} ,\qquad
\vert m_{bl^+ \nu_l} - m_{top}\vert    > 10 \: \mbox{GeV} ,\qquad
\vert m_{\bar{b}l^- \bar{\nu}_l}- m_{top}\vert    > 10 \: \mbox{GeV}.
\label{eq:cuts_vv}
\end{equation}
The latter requirement eliminates the large contribution from EW and QCD top production.

In the following we will discuss various sets of diagrams and different groups of processes, therefore,
we introduce our naming convention.
We split the amplitude $A$, for each process, as:
\begin{equation}
\label{eq:split}
A = A_h + A_H + A_0,
\end{equation}
where $A_{h/H}$ denote the set of diagrams in which a light/heavy Higgs is exchanged and $A_0$ the set
of diagrams in which no Higgs is present.  $A_{h/H}$ contain all VBS diagrams in which a $h/H$ Higgs
interacts with the vector bosons. 
They also contain a small
set of additional diagrams, e.g. Higgsstrahlung ones. These can be ignored for all practical purposes since
their contribution, with the present cut on the minimum invariant mass of the two jets which forbids them
to resonate at the mass of a weak boson, is very small. 
From time to time we will refer to the sum of subamplitudes using the notation $A_{ij}= A_{i} + A_{j}$.
A similar convention will be adopted for differential or total cross sections so that $\sigma_i$
corresponds to the appropriate integral over phase space of $\vert A_{i} \vert^2$ summed over all
contributing processes. As an example,   $\sigma_{0h}$ is obtained integrating the modulus squared of 
$A_{0h} = A_0 + A_h$, the coherent sum of the diagrams without any Higgs and those involving
the light Higgs only.

The VBS diagrams in $A_{h/H}$ can be further classified by the pair
of vector bosons which  initiate the scattering and by the final state pair.
In this paper we concentrate on
$p p \rightarrow jj l^+l^-l^{\prime +}l^{\prime -}$ and
$p p \rightarrow jj l^+\bar{\nu}_l l^{\prime -}\nu_{l^\prime}$ production so that the only instances of
VBS which appear correspond to $ZZ\rightarrow ZZ\,(Z2Z)$ and $WW\rightarrow ZZ\,(W2Z)$ for the 
$jj l^+l^-l^{\prime +}l^{\prime -}$ case and to $ZZ\rightarrow WW\,(Z2W)$ and 
$WW\rightarrow WW\,(W2W)$ for the $jj l^+\bar{\nu}_l l^{\prime -}\nu_{l^\prime}$ final state. 

The $W2Z$ and $Z2W$ sets are particularly simple because the Higgs fields
appear only in the $s$-channel. In the $Z2Z$ case
scalars are exchanged in the $s$-, $t$- and $u$-channel, while in the $W2W$ set
the Higgses contribute in the $s$- and $t$-channel.

Some of the processes contributing to 4ljj production include only the $Z2Z$ subprocess,
for instance $uc \rightarrow uc\, e^+e^- \mu^+\mu^-$; others only contain the
$W2Z$ subprocess, for instance $us \rightarrow dc\, e^+e^- \mu^+\mu^-$.
Finally
there is a class of processes, like  $ud \rightarrow ud\, e^+e^- \mu^+\mu^-$, which include both kind
of subdiagrams. They will be called $P(Z2Z)$, $P(W2Z)$ and $P(Z2Z+W2Z)$
processes respectively. 

Some processes leading to the 2l2$\nu$jj final state contain only the $Z2W$ set,
for instance $u\bar{c} \rightarrow u\bar{c}\, e^+\bar{\nu_e} \mu^-\nu_\mu$; others only contain
the $W2W$ set, like $u\bar{c} \rightarrow d\bar{s}\, e^+\bar{\nu_e} \mu^-\nu_\mu$.
A third group of reactions includes both kind of subdiagrams, for instance
$ud \rightarrow  ud\, e^+\bar{\nu_e} \mu^-\nu_\mu$.
They will be called $P(Z2W$), $P(W2W)$ and $P(Z2W+W2W)$ processes, respectively.

The 4ljj final state has a tiny branching ratio but is very clean.  The invariant mass of the
leptonic system can be measured with high precision and small background.

In the 2l2$\nu$jj final state, the two charged leptons will be
required to belong to different families and charges so that the final state can be thought of as containing
a $W^+W^-$ pair. The 2l2$\nu$jj final state has a much larger cross section. However, the invariant mass
of the $WW$ system cannot be reconstructed and it can only
be experimentally analyzed in terms of the transverse mass of the leptonic system.

In the following we will examine these reactions with the aim of clarifying the role and size of interference
effects in VBS,  disregarding their actual observability at the LHC which would require a detailed study
of all available channels and a careful assessment of reducible and irreducible backgrounds.
Some of the distributions we present are not accessible in practice but are nonetheless useful tools
for a first theoretical estimate of interference effects in different contexts.
   
\begin{figure}[tb]
\centering
\subfigure{	 
\hspace*{-3.5cm} 
\includegraphics*[width=8.3cm,height=6.2cm]{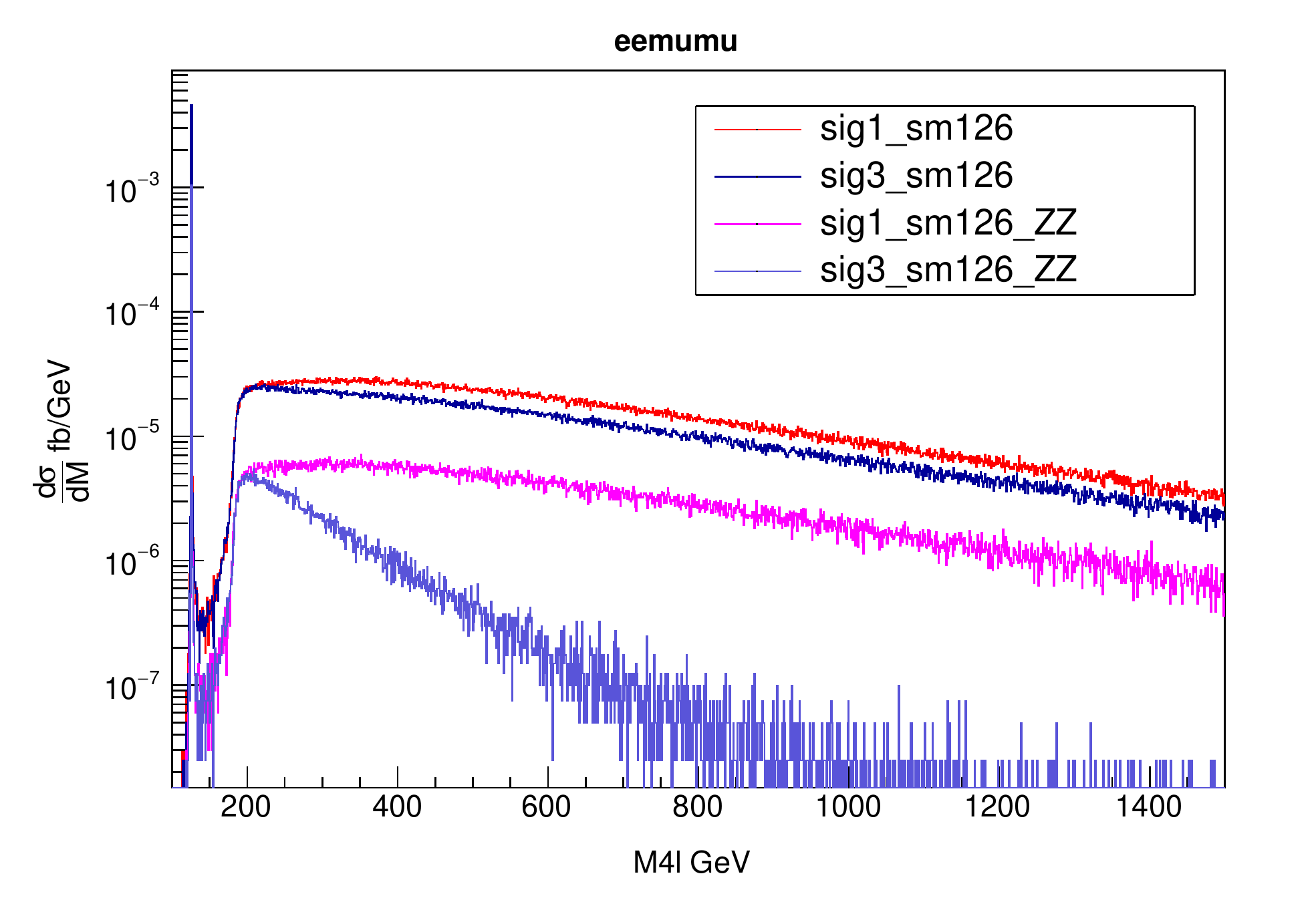}
%\hspace*{-0.7cm}
\includegraphics*[width=8.3cm,height=6.2cm]{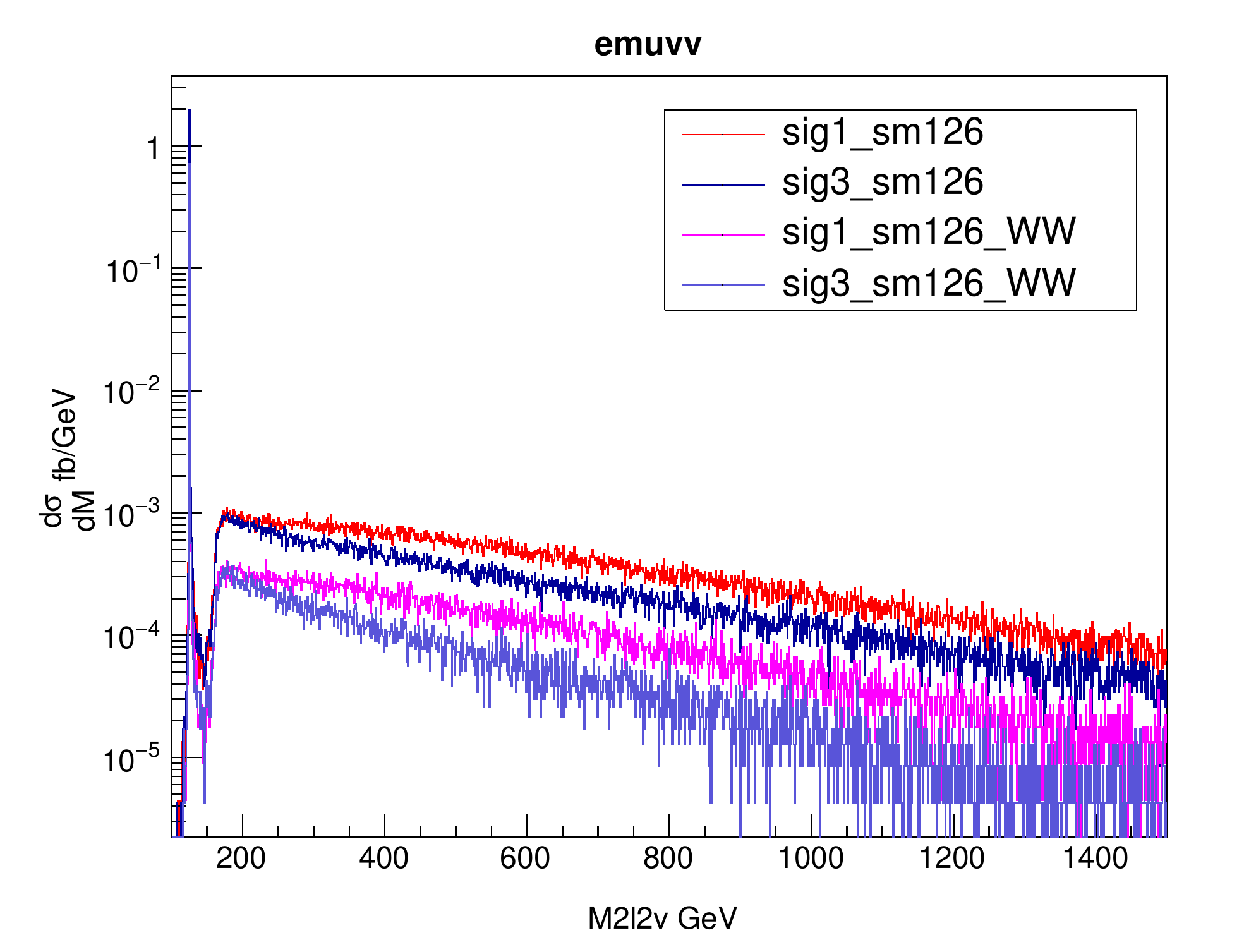}
\hspace*{-3cm}
}
\caption{
Invariant mass distribution of the four lepton system for the 4ljj final state (left)
and the 2l2$\nu$jj final state (right) in the SM.
In red and purple the mass distribution  
obtained taking into account only the diagrams with 
$s$-channel Higgs exchange and in blue and violet the result when the full set of Higgs exchange
diagrams is included.
On the left(right), the two contributions of the $P(Z2Z)$($P(W2W)$)  processes is shown separately.
}
\label{fig:fig1}
\end{figure}

We will begin our presentation with a discussion of the small set of diagrams in which VBS is mediated by
Higgs exchange. The reason for this is the Caola-Melnikov approach to
determining the Higgs width, which is based on a separation of the amplitude in a signal part,
a background part
and their interference. The three terms depend differently on the Higgs couplings, which are proportional
to the Higgs width through the peak cross section. Varying these
couplings within the experimental limits on off-shell $ZZ$ and $WW$ rates provides an upper
bound on the Higgs total width. In Run II CMS and ATLAS plan to apply this procedure to Vector Boson
Scattering and the set of diagrams in which VBS is mediated by Higgs exchange represents
the signal term.

\section{Higgs Mediated Vector Boson Scattering Signal in the SM}
\label{sec:VBS_Signal_SM}

In \fig{fig:fig1} we present results for the SM. 
On the left hand side we show in red the mass distribution for the 4ljj final state 
obtained taking into account only the diagrams with 
$s$-channel Higgs exchange and in blue the result when the full set of Higgs exchange diagrams is
included.
The contribution of the $P(Z2Z)$  processes is shown separately: in purple the result due solely to  
$s$-channel Higgs exchange and in violet the result obtained from the sum of all three channels.

On the right hand side of  \fig{fig:fig1} we show the corresponding results for the 2l2$\nu$jj final
state.  In this case, it is the contribution of the $P(W2W)$ processes which is shown separately.

We see that there is a significant difference between the curves obtained considering only $s$-channel
Higgs exchange and those obtained from the full set of scalar exchange diagrams.
This implies a conspicuous negative interference between the Higgs exchange diagrams in 
$P(Z2Z)$ and $P(W2W)$ processes. This interference is so large that it significantly modifies the
result obtained when all processes are summed, even though there are reactions which contribute
substantially to the total which are not affected at all by these effects like $P(W2Z)$ and $P(Z2W)$
processes and others, the $P(Z2W+W2W)$ and  $P(Z2Z+W2Z)$ groups, which are affected only partially. 

Large cancellations in $P(Z2Z)$ processes are expected.
On shell $ZZ \rightarrow ZZ$ scattering is zero in the absence of the Higgs and therefore does not
violate
unitarity at high energy. As a consequence the corresponding Higgs diagrams, each of which grows
as the invariant mass squared of the process, must combine in such a way that their sum is actually 
asymptotically finite. 
At large energy, the longitudinal 
polarization vector of a Z boson of momentum $p^\mu$ can be identified with
$p^\mu/M_Z$ and the sum of the three Feynman diagrams describing the scattering behaves as
$s^2/s+t^2/t+u^2/u=s+t+u\approx 0$. It is however surprising that  the cancellation grows very
rapidly, above threshold, with the mass of the ZZ pair and becomes substantial already at moderate
invariant masses.
For $M_{ZZ} = $ 500 GeV the square of the
three Higgs exchange diagrams is an order of magnitude smaller than the result obtained from the
$s$-channel exchange alone.
The same cancellation takes place in the amplitude of the $P(Z2Z+W2Z)$ processes, while the
$P(W2Z)$ sector is unaffected.
In the sum of  all processes the interference decreases the SM result for 
$s$-channel Higgs exchange by about 25\%.

Interference effects are present also in $P(W2W)$ processes, as shown in the right hand side of
\fig{fig:fig1}. They are less prominent than in the  $P(Z2Z)$ case.
The same cancellation takes place in the amplitude of the $P(Z2W+W2W)$ processes, while the 
$P(Z2W)$ sector is unaffected.
Summing all processes, the difference between the result obtained
from the single  $s$-channel exchange diagram (red) and the full set (blue) is larger than for 4ljj
production
because WW initiated scatterings are more frequent than ZZ ones for the 2l2$\nu$jj final state.
The interference decreases the SM result for $s$-channel Higgs exchange by about 30\%.
The on shell reaction $W^+W^- \rightarrow W^+W^-$ violates unitarity in a Higgsless theory 
when
the $W$'s are longitudinally polarized. Therefore Higgs exchange diagrams are necessary to 
restore unitarity and the cancellation can only be partial.
There is no $u$-channel exchange, so, at large energy, the two diagrams behave as
$t^2/t+s^2/s=t+s\approx -u$. 

These results imply that, when producing Monte Carlo templates for the analysis of off shell Higgs
production, it is mandatory to include the full set of Higgs exchange diagrams. This is in agreement with
the Caola-Melnikov method which isolates terms in the amplitude which are proportional to the same
power of the Higgs couplings. As a consequence all Higgs exchange diagrams need to be taken as a unit,
regardless of the channel in which the exhange takes place.
A production times decay approach is clearly inadequate to describe the off shell
Higgs contribution. QCD radiative corrections in VBF are small.
They are crucial in reducing the scale dependence of the predictions to the 5-10\% level.
NNLO corrections bring the uncertainty down to about 2\%. When aiming for such an accuracy,
interference effects, which have a comparable if not larger impact, cannot be ignored.

\section{Higgs Mediated Vector Boson Scattering Signal in the 1HSM}
\label{sec:VBS_Signal_1HSM}

\begin{figure}[tb]
\centering
\subfigure{	 
\hspace*{-3.5cm} 
\includegraphics*[width=8.3cm,height=6.2cm]{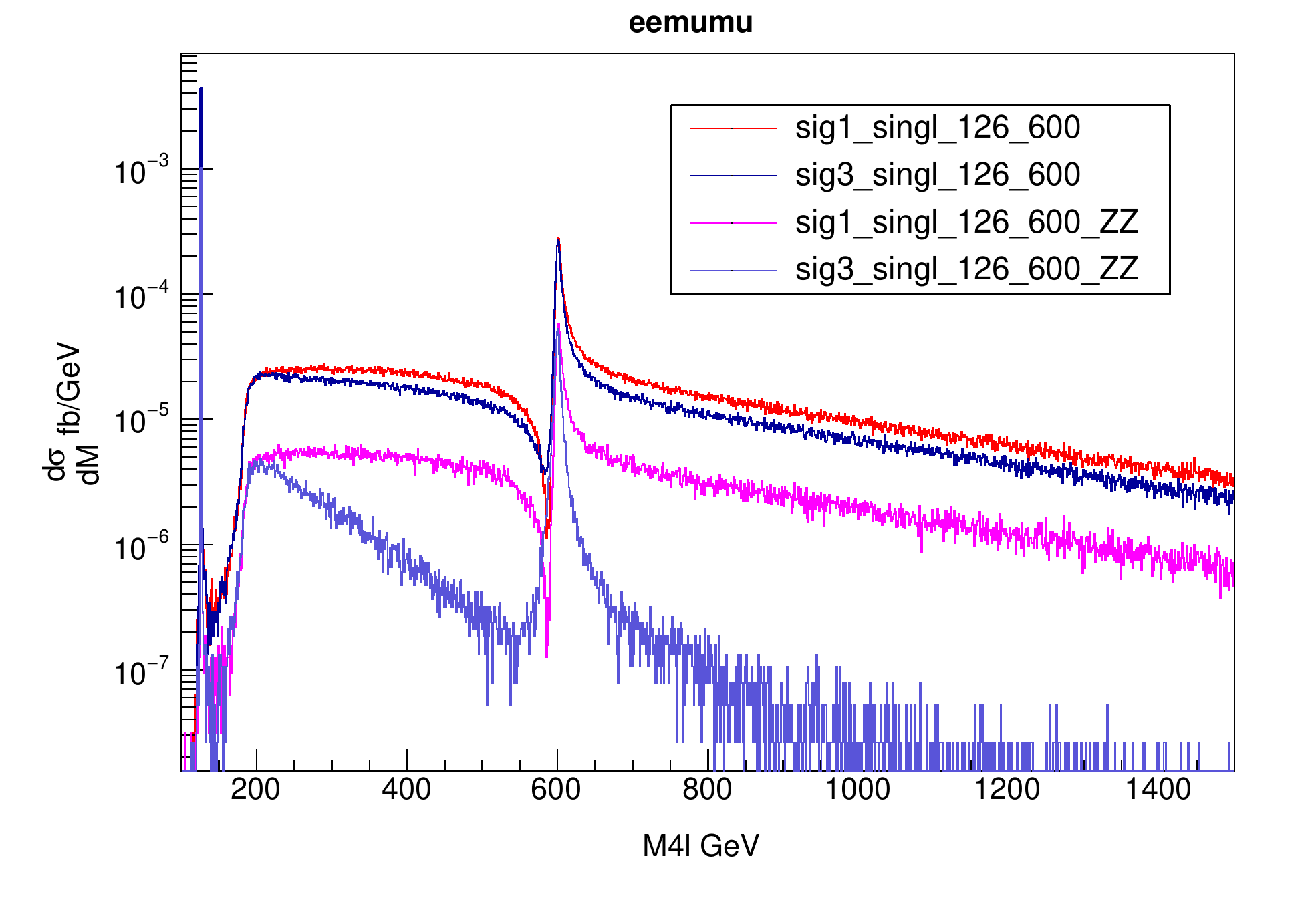}
%\hspace*{-0.7cm}
\includegraphics*[width=8.3cm,height=6.2cm]{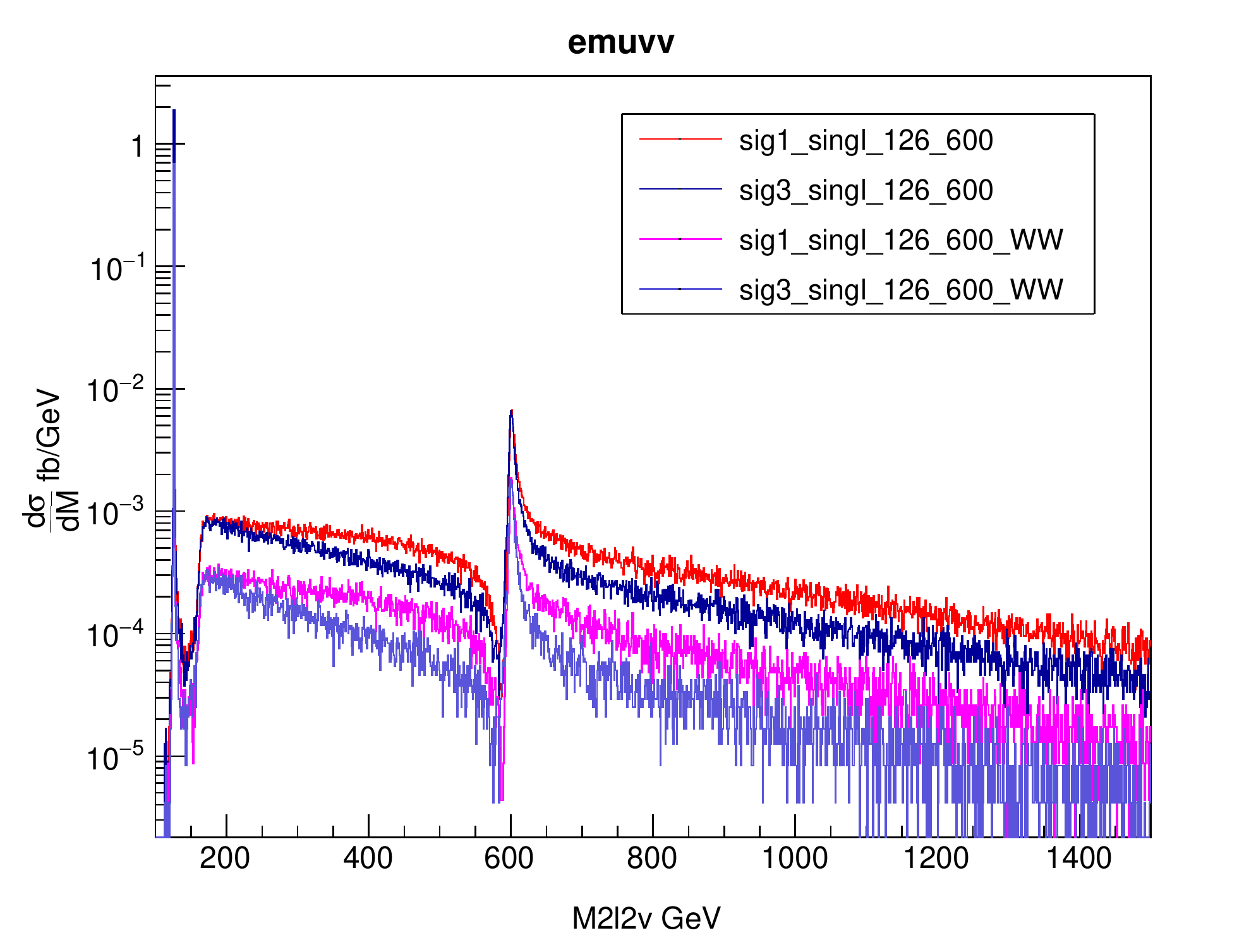}
\hspace*{-3cm}
}
\caption{
Invariant mass distribution of the four lepton system for the 4ljj final state (left)
and the 2l2$\nu$jj final state (right) in the 1HSM with $M_H$= 600 GeV and $s_\alpha=0.2$,
$\tan\beta=0.3$.
In red and purple the mass distribution  
obtained taking into account only the diagrams with 
$s$-channel Higgs exchange and in blue and violet the result when the full set of Higgs exchange
diagrams is included.
On the left(right), the two contributions of the $P(Z2Z)$($P(W2W)$)  processes is shown separately.
}
\label{fig:fig2}
\end{figure}

\begin{figure}[tb]
\centering
\subfigure{	 
\hspace*{-3.5cm} 
\includegraphics*[width=8.3cm,height=6.2cm]{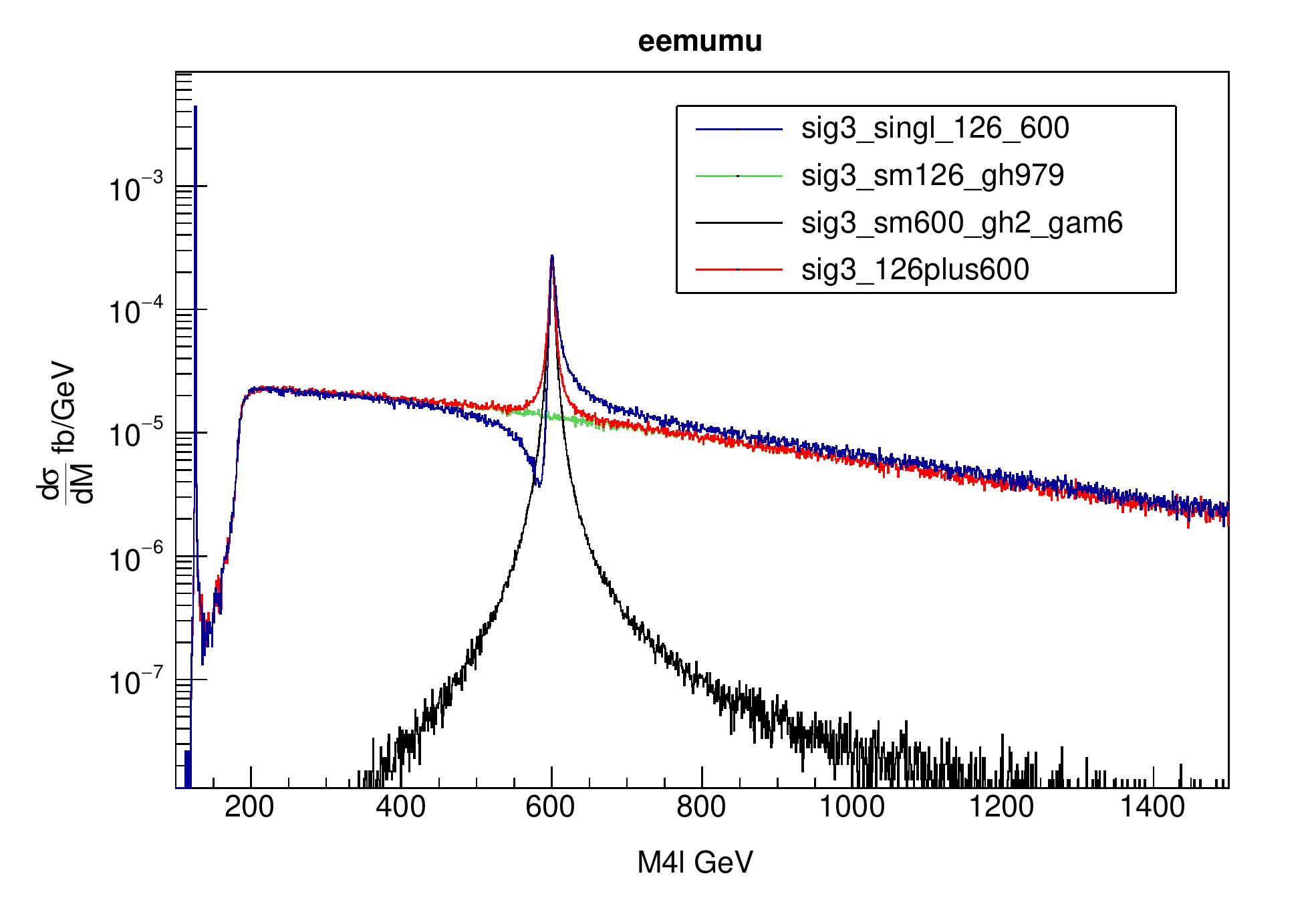}
\hspace*{-0.7cm}
\includegraphics*[width=8.3cm,height=6.2cm]{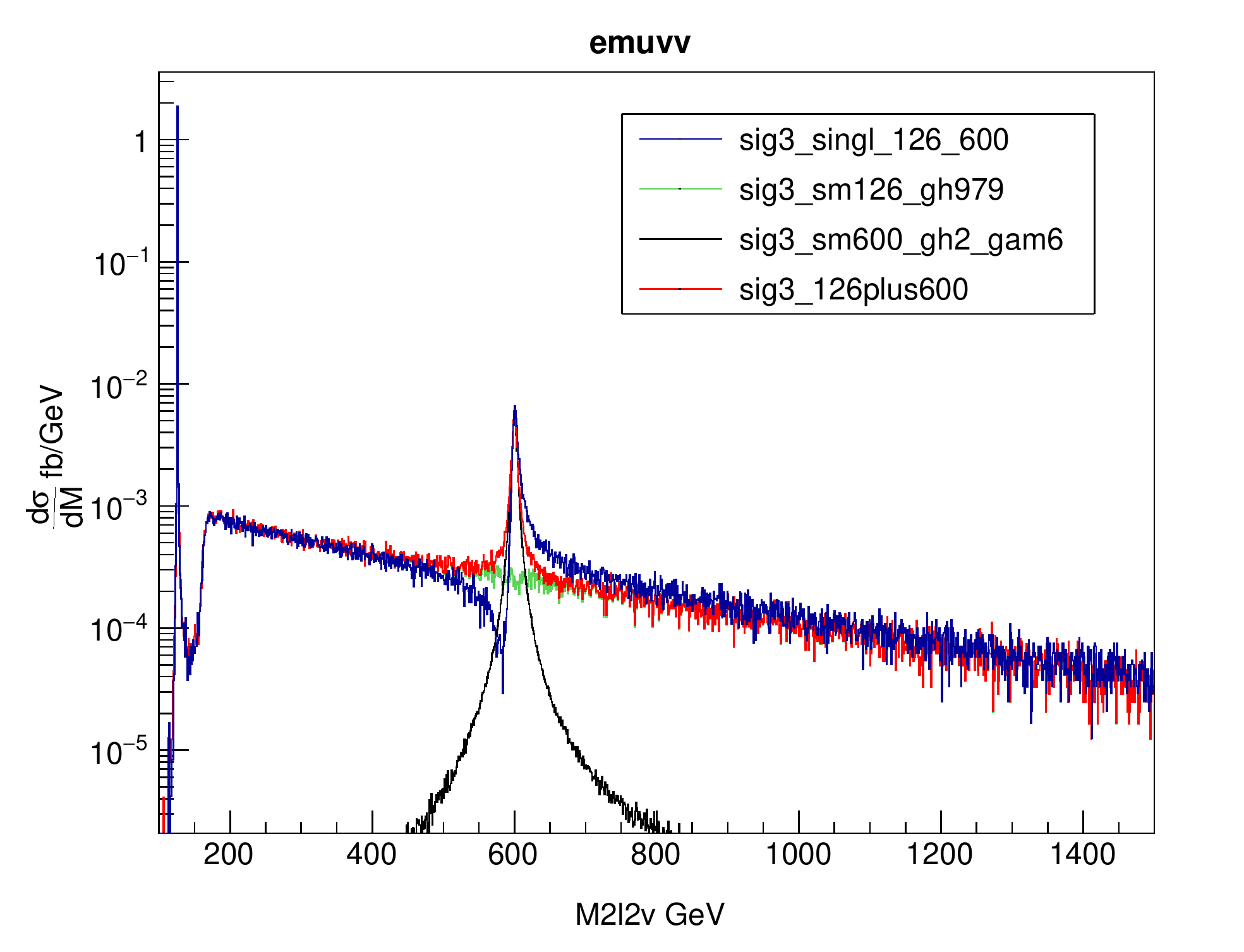}
\hspace*{-3cm}
}
\caption{
Invariant mass distribution of the four lepton system for the 4ljj final state
in the 1HSM with $M_H$= 600 GeV and $s_\alpha=0.2$, $\tan\beta=0.3$.
In green and black the mass distribution  
obtained taking into account the full set of Higgs exchange diagrams
for Higgs masses of 125 and 600 GeV respectively. In red the incoherent sum of the two contributions.
In blue the result of all Higgs diagrams in the 1HSM. 
}
\label{fig:fig3}
\end{figure}

We now turn to the 1HSM. We present results for selected values of  $M_H,\ s_\alpha$
and $\tan\beta$ but our conclusions are fairly independent of the choice of parameters.
In \fig{fig:fig2} we show a number of four lepton mass distributions,
for the 4ljj final state on the left and the 2l2$\nu$jj final state on the right, for $M_H=600$ GeV,
$s_\alpha=0.2$ and $\tan\beta=0.3$.
The colors of the histograms in \fig{fig:fig2} follow the convention of 
\fig{fig:fig1}. The red and purple lines refer to pure $s$-channel exchange.  The red one relates to the
sum of all processes while the purple one to $P(Z2Z)$ processes (left) and $P(W2W)$ ones (right), only.
The blue and violet lines correspond to the sum of all Higgs exchange diagrams.
The pattern and size of interference effects among different sets of Higgs exchange diagrams are similar
to those in the SM. 
In addition, all curves in \fig{fig:fig2}, in the region around 600 GeV, show 
an interference pattern between the light and heavy Higgs similar to one
present in the GGF case \cite{Englert:2014ffa,Maina:2015ela,Kauer:2015hia,Englert:2015zra}.
The inclusion of the full set of Higgs exchange diagrams decreases the size of the pure 
$s$-channel exchange amplitude over the whole energy range, as in the SM case,
with the exception of a small region below the heavy Higgs mass where the interference between the two
scalars dominate. It also significantly affects the interference pattern in the neighborhood of $M_H$.

In \fig{fig:fig3} we compare the invariant mass distribution of the four lepton system for the 4ljj (left)
and 2l2$\nu$jj (right) final state
obtained taking into account the full set of Higgs exchange diagrams in the 1HSM (blue) with the
incoherent sum (red) of the Higgs exchange diagrams for Higgs masses of 125 and 600 GeV.
The individual contributions of the two Higgs are shown in green and black respectively.  
The difference between the blue curve and the black and red ones illustrates the deformation of the Breit
Wigner distribution induced by interference effects.
They are negative in the region below $M_H$ and positive above the heavy Higgs resonance as
demonstraded by the comparison of the blue and red histograms.
Effects are even larger if only the $s$-channel exchange
is taken into account but from now on we only consider the
full set of Higgs exchange diagrams which, even though not gauge invariant and therefore not physically
observable, provides a better description of the Higgs contribution in the off shell region. 

Clearly, this interference between different Higgs fields is not a peculiarity of the Singlet Model. It 
will indeed occur in any theory with  multiple scalars which couple to the same set of elementary
particles, albeit possibly with different strengths.

\section{Full processes}
\label{sec:full}

\begin{figure}[tb]
\centering
\subfigure{	 
\hspace*{-3.5cm} 
\includegraphics*[width=8.3cm,height=6.2cm]{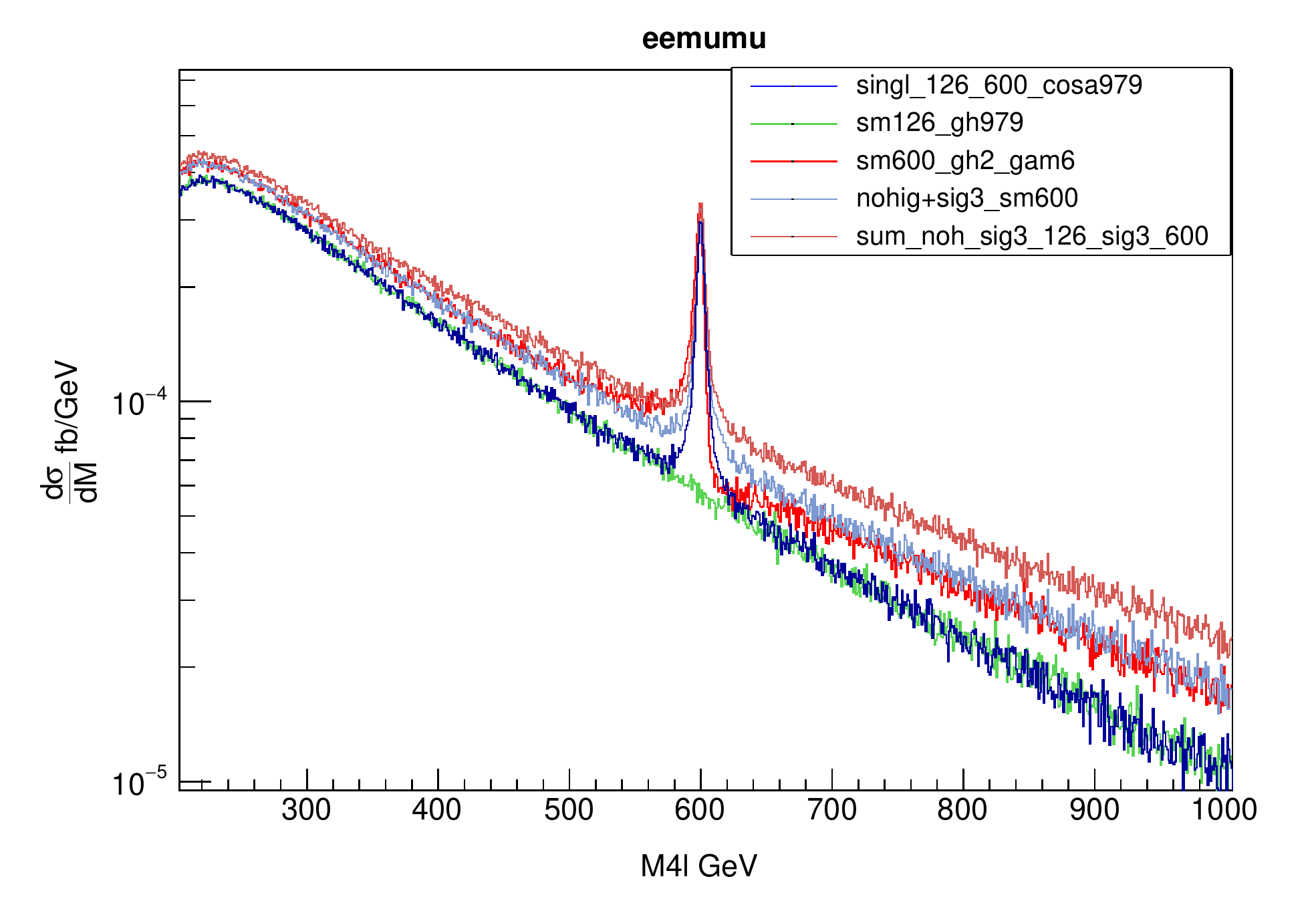}
%\hspace*{-0.7cm}
\includegraphics*[width=8.3cm,height=6.2cm]{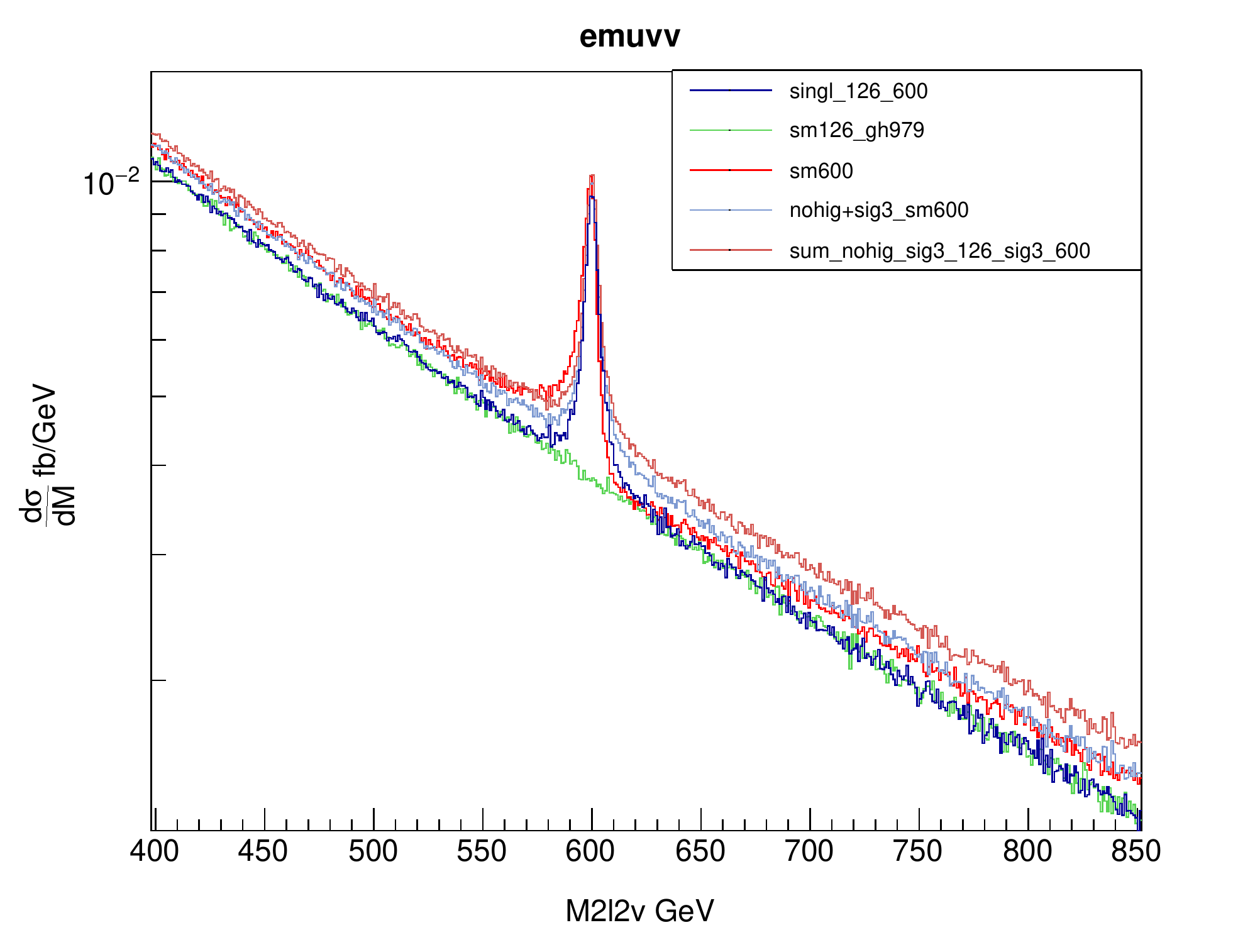}
\hspace*{-3cm}
}
\caption{
In blue, the invariant mass distribution of the four lepton system for the 4ljj final state (left)
and the 2l2$\nu$jj final state (right) in the 1HSM with $M_H$= 600 GeV and $s_\alpha=0.2$.
The other curves are different approximations as detailed in the main text.
}
\label{fig:fig4}
\end{figure}

After our presentation of the interplay of the different sets of Higgs exchange diagrams, 
we move to the discussion of the actual cross section for the production of a Singlet Model heavy Higgs at
the LHC.
The plot on the left hand side of \fig{fig:fig4} shows the prediction for 4ljj production in the 1HSM (blue)
with $M_H$= 600 GeV and $s_\alpha=0.2$.
Charged leptons satisfy the requirements in 
\eqn{eq:cuts_lep} while jets pass the cuts in \eqn{eq:cuts_jet}.
The 1HSM exact result is compared with different approximations.
The green histograms is the light Higgs plus no-Higgs contribution, $d\sigma_{0h}/dM$;
the red one refers to $d\sigma_{0H}/dM$; 
the gray one to $d\sigma_{0}/dM + d\sigma_{H}/dM$
and the brown one to $d\sigma_{0}/dM + d\sigma_{h}/dM + d\sigma_{H}/dM$.
On the right hand side of \fig{fig:fig4} the corresponding curves for the 2l2$\nu$jj final state are
dispayed.
None of the approximations in \fig{fig:fig4} approaches the exact result better than about 20\% in the
region around the heavy scalar peak
and they obviously fare even worse at large $M_{4l}$, with the exception of the green curve which misses
only the heavy Higgs subamplitude, which is proportional to $s_\alpha^2$ and numerically small
in this energy range and outside the peak region, though necessary for unitarity.
Clearly, neglecting any part of an amplitude requires a great deal of attention and a careful estimate of the
resulting discrepancy.

\begin{figure}[tb]
\centering
\subfigure{	 
\hspace*{-3.5cm} 
\includegraphics*[width=8.3cm,height=6.2cm]{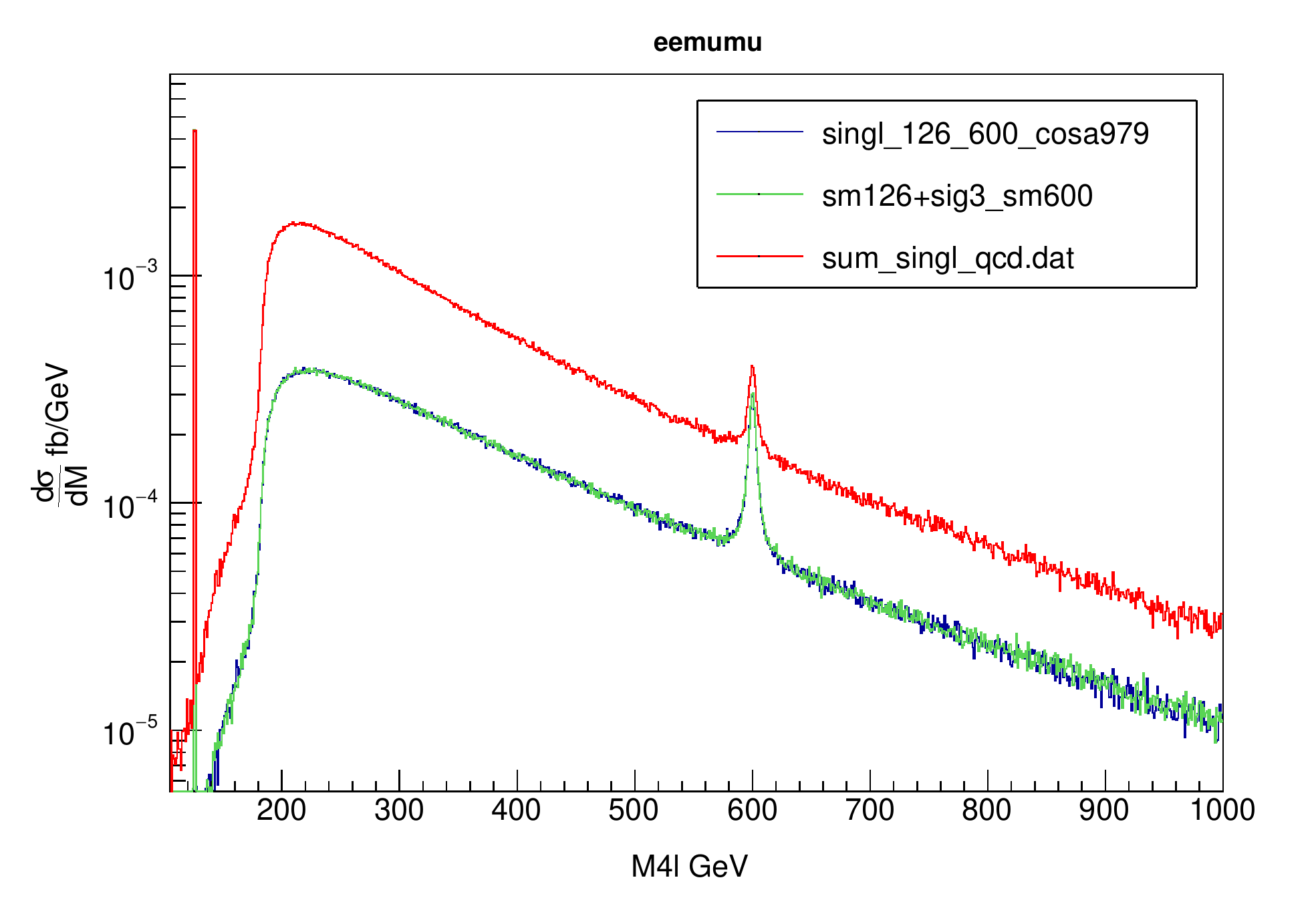}
%\hspace*{-0.7cm}
\includegraphics*[width=8.3cm,height=6.2cm]{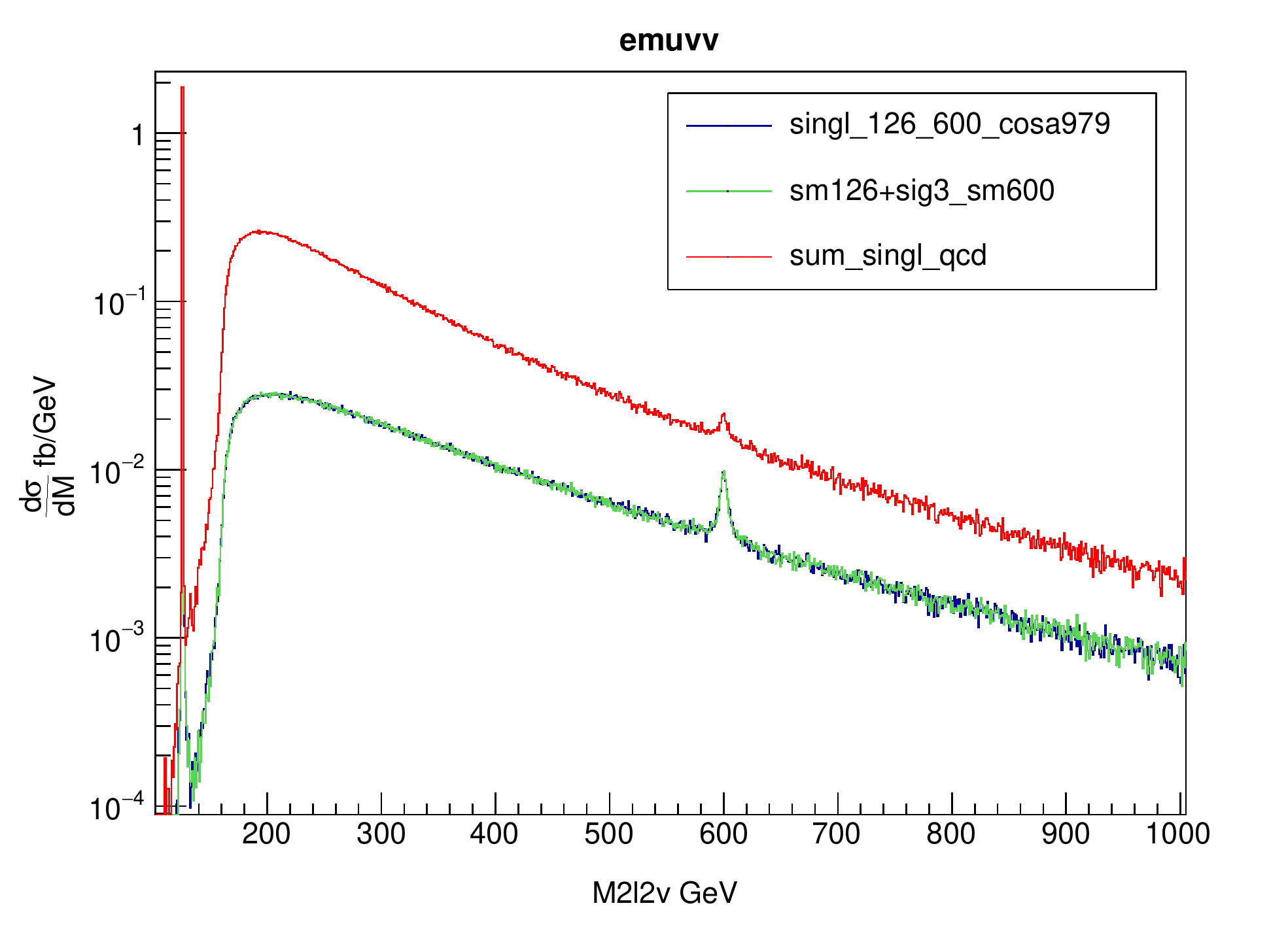}
\hspace*{-3cm}
}
\caption{
Invariant mass distribution of the four lepton system for the 4ljj final state (left)
and the 2l2$\nu$jj final state (right) in the 1HSM with $M_H$= 600 GeV and $s_\alpha=0.2$.
The blue histogram is the exact 1HSM result. The green line refers to 
$d\sigma_{0h}/dM +  d\sigma_{H}/dM$.
The red curve is the sum of the 1HSM result and of the QCD contribution at $\ordQCD$.
}
\label{fig:fig5}
\end{figure}

\begin{figure}[tb]
\centering
\includegraphics*[width=8.3cm,height=6.2cm]{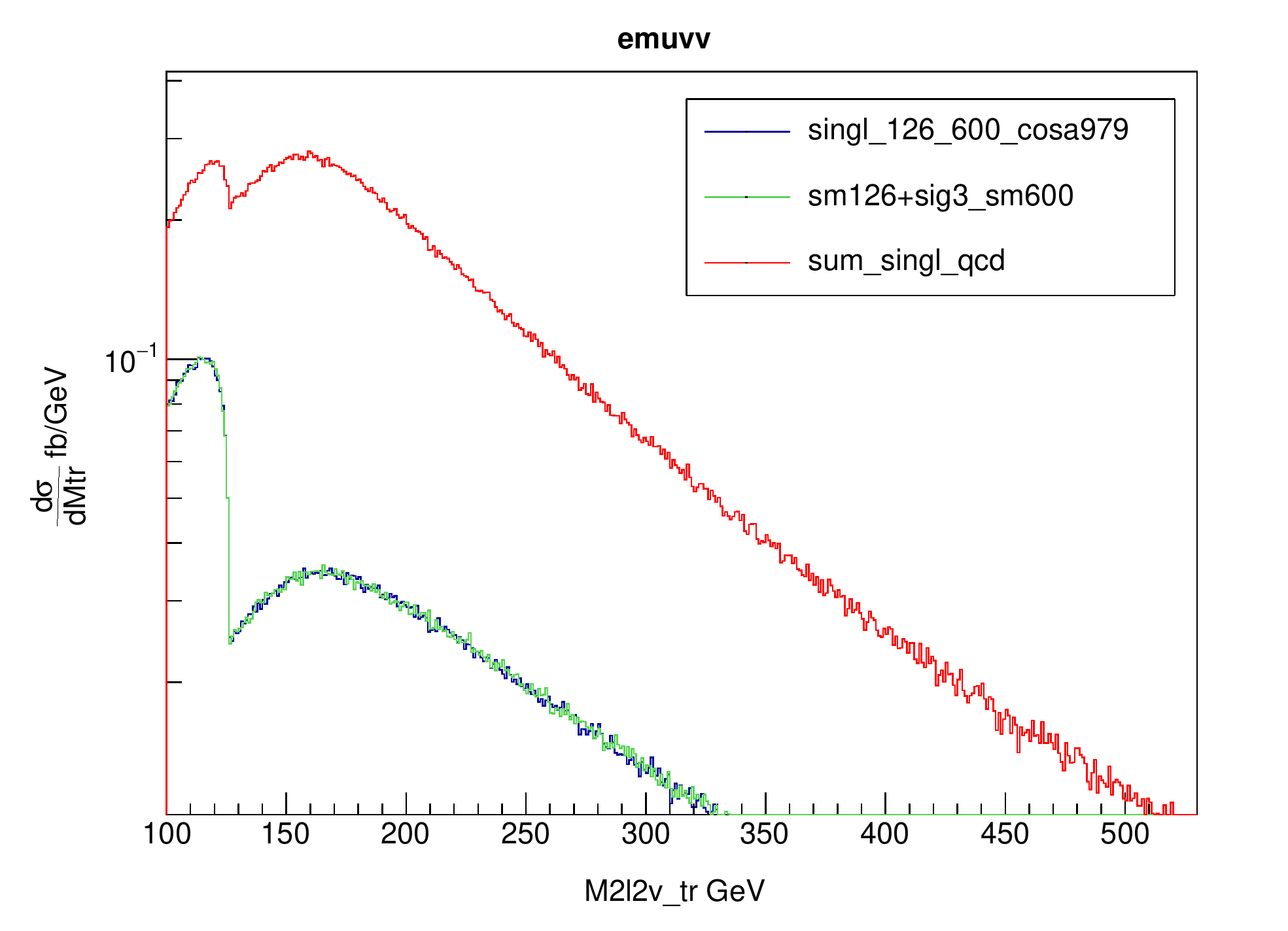}
\caption{
Transverse mass distribution of the four lepton system for the  2l2$\nu$jj final state in the 1HSM
with $M_H$= 600 GeV and $s_\alpha=0.2$.
The blue histogram is the exact 1HSM result. The green line refers to 
$d\sigma_{0h}/dM +  d\sigma_{H}/dM$.
The red curve is the sum of the 1HSM result and of the QCD contribution at $\ordQCD$.
}
\label{fig:fig6}
\end{figure}

\begin{figure}[tb]
\centering
\subfigure{	 
\hspace*{-3.5cm} 
\includegraphics*[width=8.3cm,height=6.2cm]{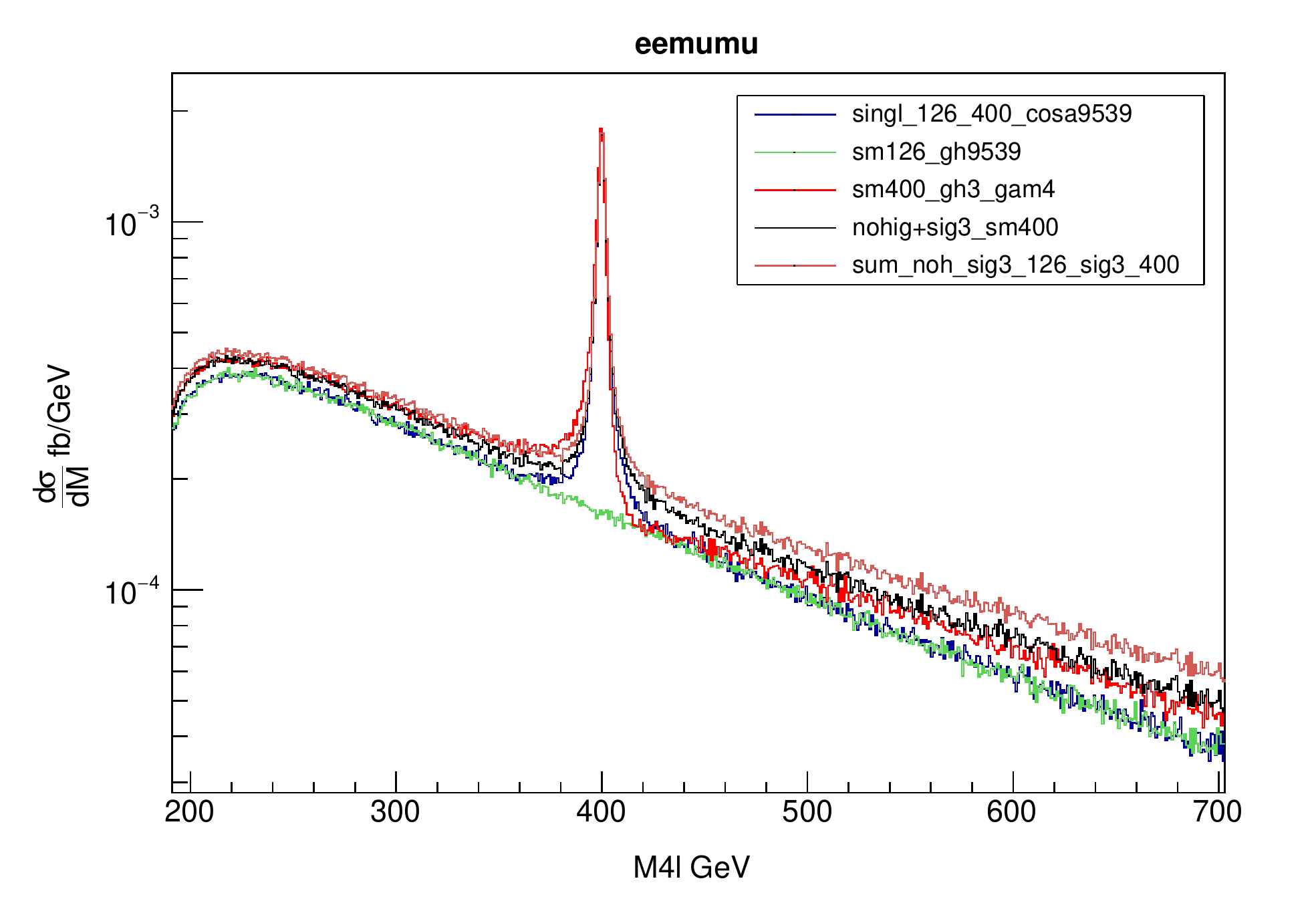}
%\hspace*{-0.7cm}
\includegraphics*[width=8.3cm,height=6.2cm]{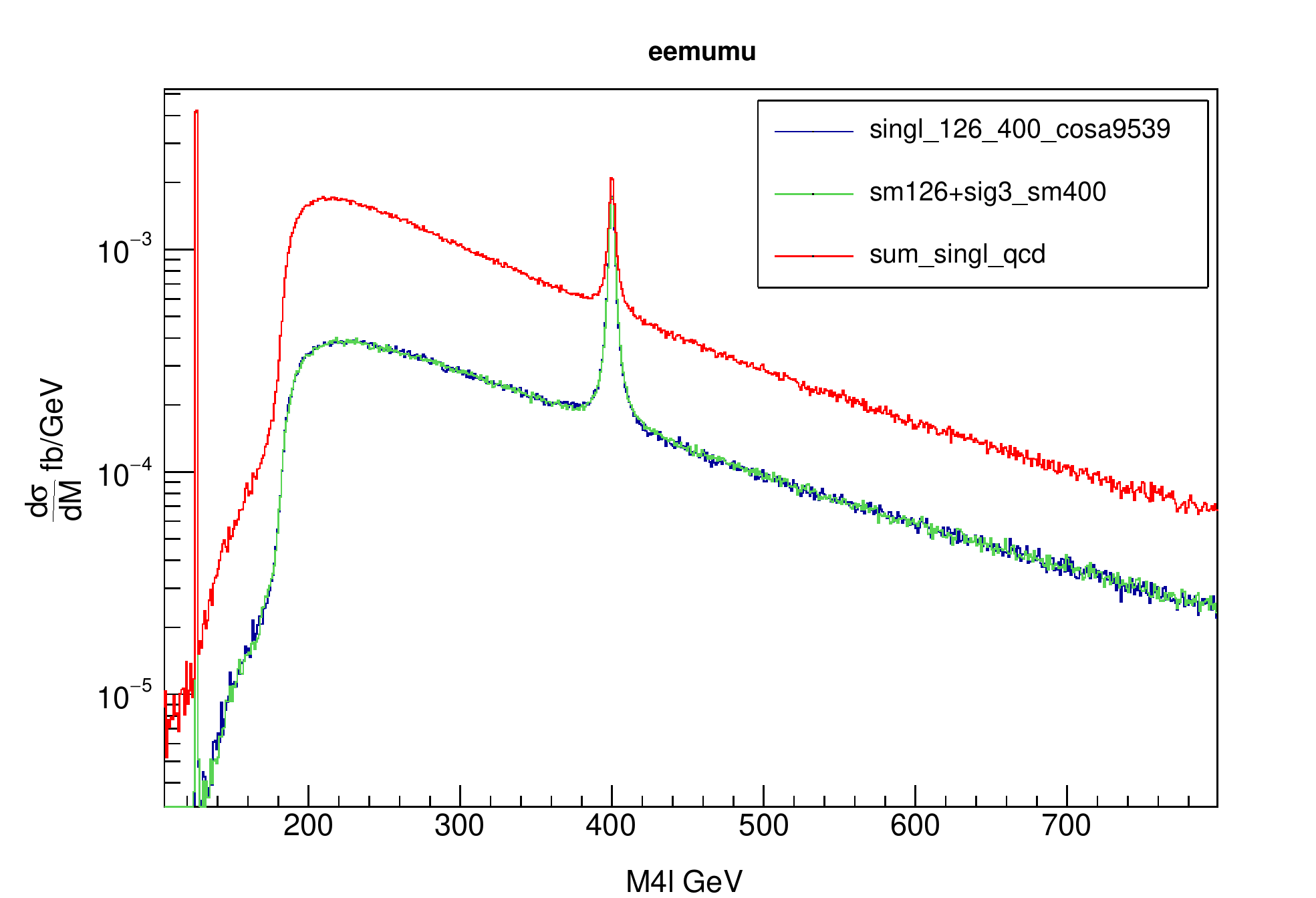}
\hspace*{-3cm}
}
\caption{
Invariant mass distribution of the four lepton system for the 4ljj final state in the 1HSM with 
$M_H$= 400 GeV, $s_\alpha=0.3$ and $\tan\beta=1.0$.
On the left, the 1HSM result, in blue, is compared
with different approximations as detailed in the main text. 
On the right the exact result is compared with $d\sigma_{0h}/dM+d\sigma_{M}/dM$,in green. 
The red curve is the sum of the 1HSM result and of the QCD contribution at $\ordQCD$.
}
\label{fig:fig7}
\end{figure}

There is however a combination of subamplitudes which provides a good approximation to the exact
result.
In \fig{fig:fig5} the prediction for 4ljj/2l2$\nu$jj production in the 1HSM, in blue, is compared
with the curve, in green, obtained from the incoherent sum of $d\sigma_{0h}/dM$ and 
$d\sigma_{H}/dM$, both of them computed with with 1HSM couplings and widths.
The two histograms agree remarkably well over the full mass range.
This is particularly meaningful in the region of the heavy Higgs peak where  $A_H$ is large:
it implies that the interference terms of the heavy Higgs diagrams with $A_h$ and $A_0$ cancel each
other to a large degree.
For comparison, we also show in red the sum of the full $\ordEW$ result discussed above and
of the QCD contribution at $\ordQCD$.  The cross section is a factor of about three larger than the EW
result.

As mentioned before, the invariant mass of the $W$ boson pair is not measurable, therefore
in \fig{fig:fig6} we show the transverse mass distribution for the 2l2$\nu$jj final state. 
The transverse mass is defined as:
\begin{equation}
\left( M_T^{WW} \right)^2 = \left( E_{T,ll} + E_{T,miss} \right)^2 - \lvert \vec{p}_{T,ll} + \vec{E}_{T,miss}
\rvert^2, 
\label{t-mass}
\end{equation}
where $E_{T,ll} = \sqrt{\left( \vec{p}_{T,ll} \right)^2+ M^2_{ll} }$.
The heavy Higgs
peak has been completely washed out, as expected. 
Also in this case, the sum $\sigma_{0h} + \sigma_H$ describes very well the exact distribution.
Clearly the fully leptonic decay of the $WW$ pair can
only be considered as a case study. In order to employ the $W^+W^-jj$ channel in the search for
additional heavy scalars it will be necessary to consider the semileptonic decays.   

As a check of the dependence of the effects discussed above on the heavy Higgs mass,
in \fig{fig:fig7} we show some results for 4ljj production in the 1HSM with $M_H$= 400 GeV, 
$s_\alpha=0.3$ and $\tan\beta=1.0$.
On the left, the full result, in blue, is compared with different combinations of subamplitudes.
The green histograms is the light Higgs plus no-Higgs contribution, $d\sigma_{0h}/dM$;
the red one refers to $d\sigma_{0H}/dM$; 
the black one to $d\sigma_{0}/dM + d\sigma_{H}/dM$
and the brown one to $d\sigma_{0}/dM + d\sigma_{h}/dM + d\sigma_{H}/dM$.
Again, none of these approximations describe satisfactoraly the region around the heavy scalar peak.
All of them, with the exception of the green curve, lack terms which are crucial for the restoration of
unitarity, and progressively diverge from the exact result as the four lepton mass increases.
On the right the exact result is compared with $d\sigma_{0h}/dM+d\sigma_{M}/dM$. 
The agreement between two curves is impressive.
In red we show the sum of the full $\ordEW$ result and
of the QCD contribution at $\ordQCD$.

\section{Cancellation of the heavy Higgs interferences}
\label{sec:Cancellation}

\begin{figure}[tb]
\centering
\subfigure{	 
\hspace*{-3.5cm} 
\includegraphics*[width=8.3cm,height=6.2cm]{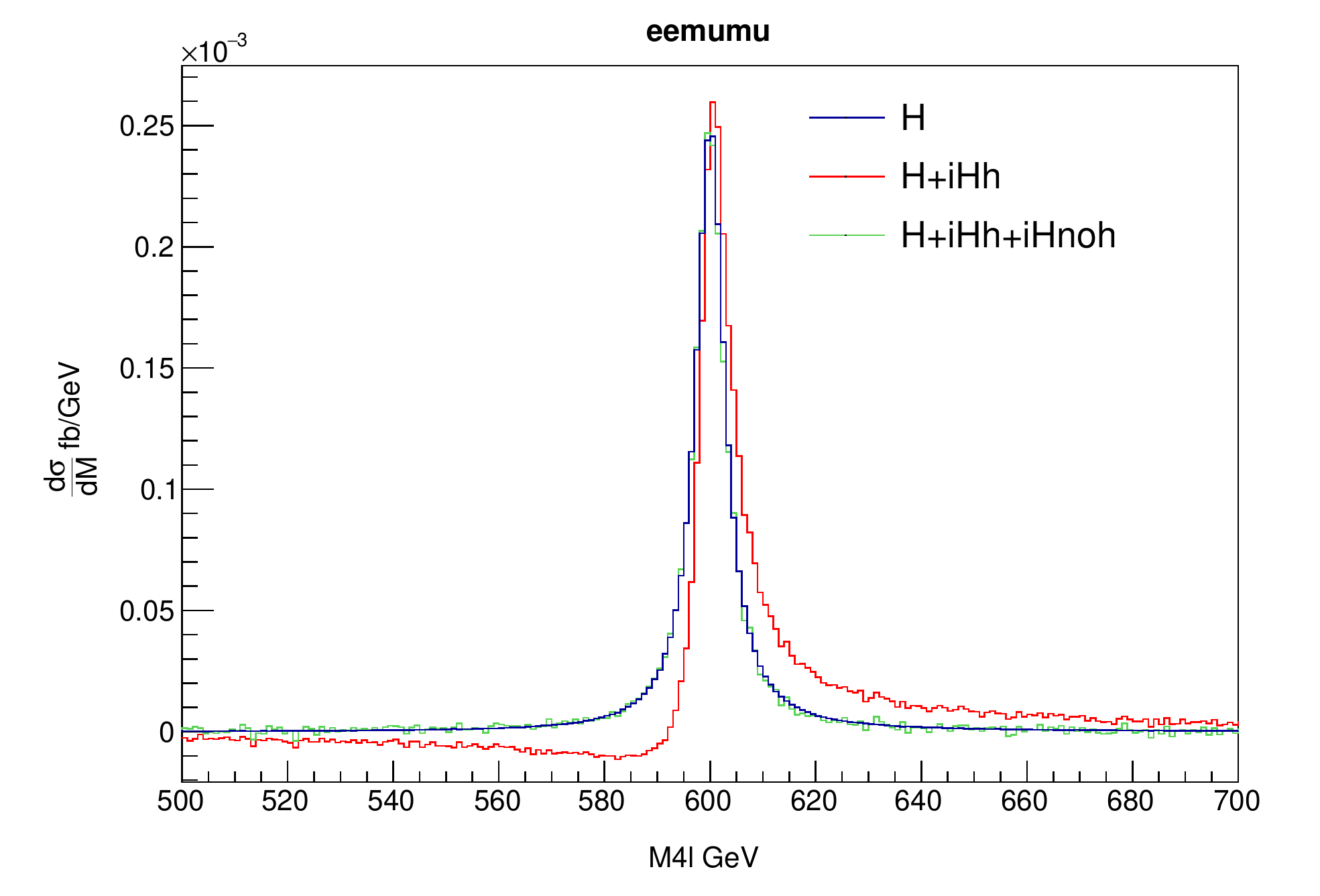}
%\hspace*{-0.7cm}
\includegraphics*[width=8.3cm,height=6.2cm]{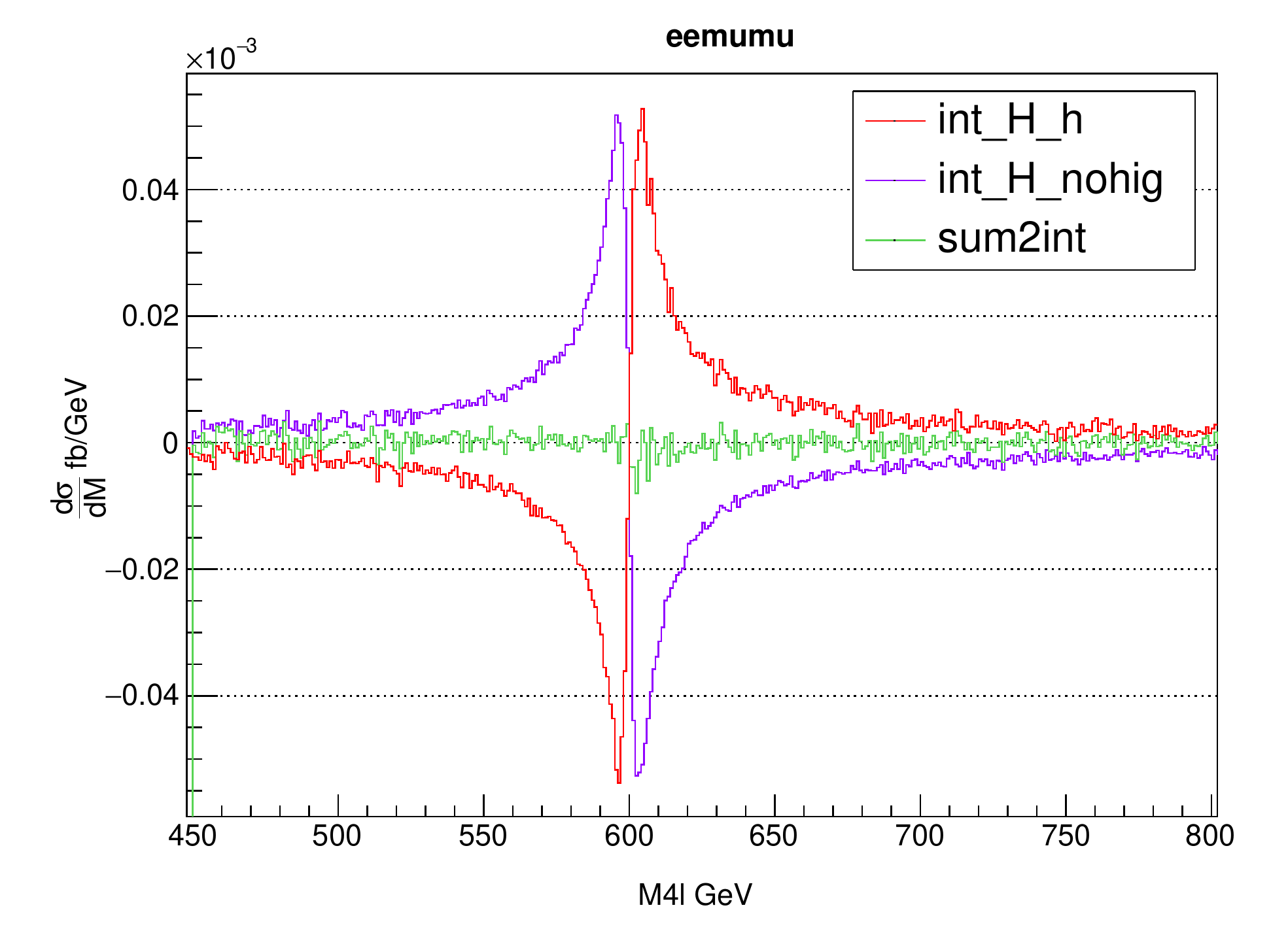}
\hspace*{-3cm}
}
\subfigure{	 
\hspace*{-3.5cm} 
\includegraphics*[width=8.3cm,height=6.2cm]{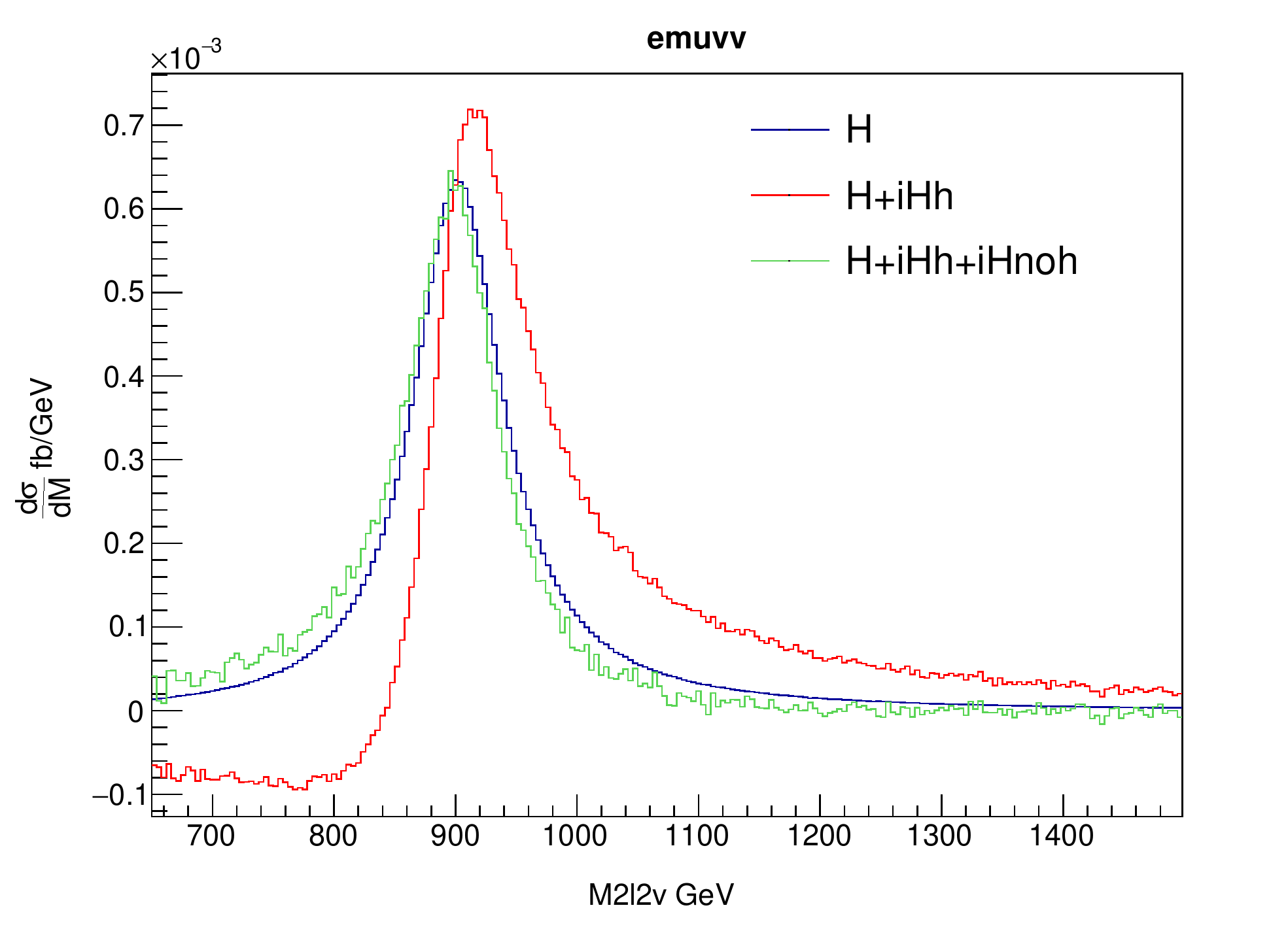}
%\hspace*{-0.7cm}
\includegraphics*[width=8.3cm,height=6.2cm]{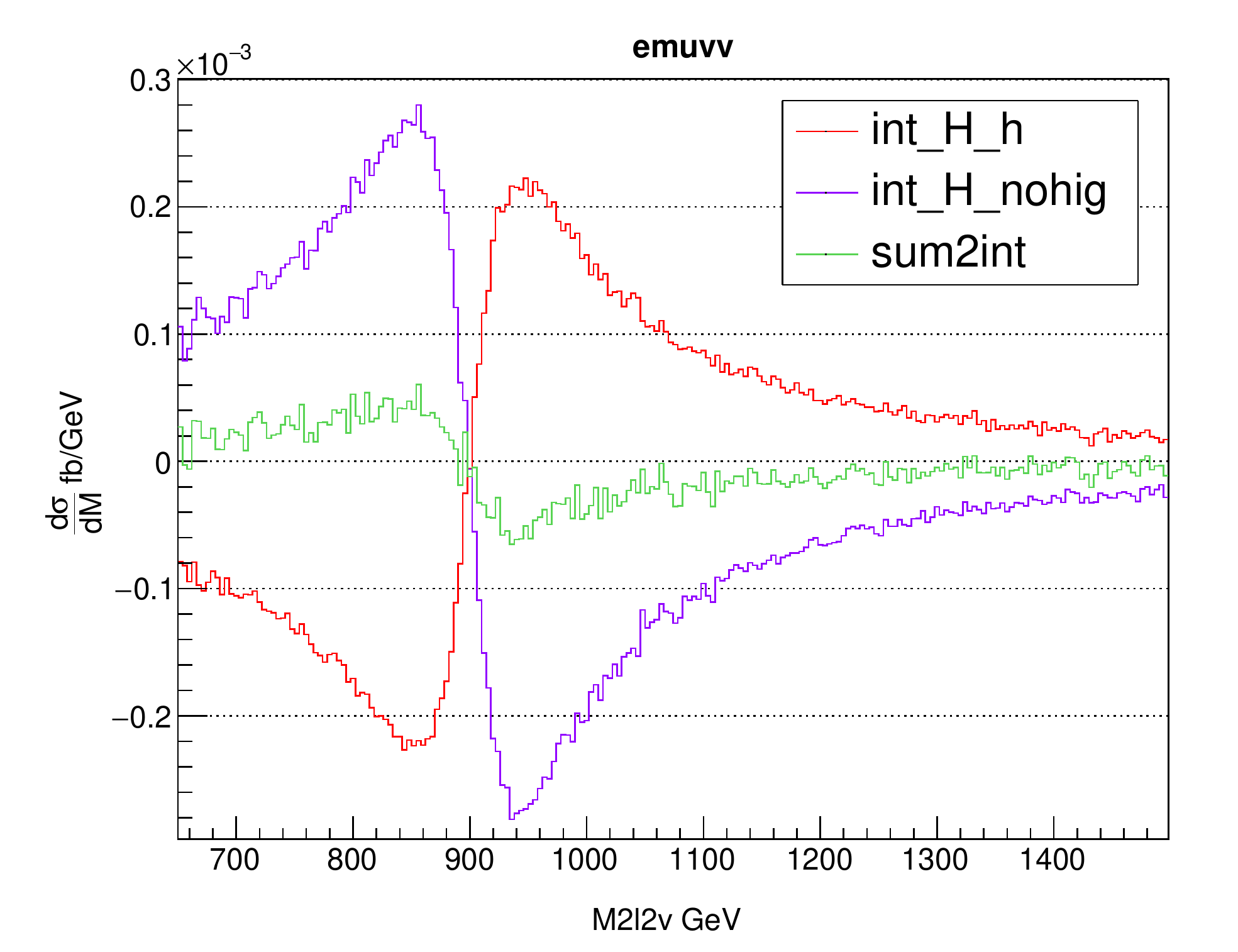}
\hspace*{-3cm}
}
\caption{
In the upper row, the invariant mass distribution of the four lepton system for the 4l final state in
the 1HSM with $M_H$= 600 GeV and $s_\alpha=0.2$.
In the lower row the corresponding plots for the 2l2$\nu$jj final state with $M_H$= 900 GeV and 
$s_\alpha=0.4$.
On the left we show $d\sigma_{H}/dM$ (blue), $d\sigma_{H}/dM+dI_{hH}/dM$ (red)
and  $d\sigma_{H}/dM+dI_{hH}/dM+dI_{0H}/dM$ (green).
On the right we show $dI_{hH}/dM$ (red),
$dI_{0H}/dM$ (violet)
and  $dI_{hH}/dM+dI_{0H}/dM$ (green).
}
\label{fig:fig8}
\end{figure}

It is noteworthy that the interference terms of the heavy Higgs diagrams with $A_h$ and $A_0$ cancel
each other almost exactly for different ranges of invariant mass of the final state vector boson pair and
different small amount of mixing between the light and heavy Higgs. 
The interference corresponds to the real part of $A_H^\ast\times (A_0+A_h)$. 
Since $A_h \propto  c^2_\alpha$,  the cancellation cannot take place for arbitrary values of of the mixing
angle $\alpha$.

In order to investigate further this phenomenon, in \fig{fig:fig8} we isolate the interference term for
different choices of parameters. 
In the upper row, the invariant mass distribution of the four lepton system for the 4l final state in
the 1HSM with $M_H$= 600 GeV and $s_\alpha=0.2$.
In the lower row the corresponding plots for the 2l2$\nu$jj final state with $M_H$= 900 GeV and 
$s_\alpha=0.4$,  a rather extreme case in view of the allowed parameter space.
Defining $I_{ij}$ as the integrated interference between $A_i$ and $A_j$,
on the right we show $d\sigma_{hH}/dM-d\sigma_{h}/dM-d\sigma_{H}/dM=dI_{hH}/dM$ (red),
$d\sigma_{H0}/dM-d\sigma_{0}/dM-d\sigma_{H}/dM=dI_{0H}/dM$ (violet)
and  $dI_{hH}/dM+dI_{0H}/dM$ (green).
On the left we show $d\sigma_{H}/dM$ in blue, $d\sigma_{H}/dM+dI_{hH}/dM$ (red)
and  $d\sigma_{H}/dM+dI_{hH}/dM+dI_{0H}/dM$ (green).

The plot in the upper left corner shows how, for $M_H$= 600 GeV and $s_\alpha=0.2$, the interference
between the heavy and the light Higgs deforms the Breit-Wigner distribution of the heavy scalar and how
the inclusion of the interference between the heavy Higgs and the subamplitude without any Higgs
pratically eliminates the deformation. The plot on the top right displays the two interferences and their
sum, which is zero within statistical uncertainty.
The two plots in the lower part provide the same information for the  2l2$\nu$jj final state with $M_H$=
900 GeV and $s_\alpha=0.4$. Since now $c^2_\alpha=0.84$ is larger than in the previous example,
the interference between the heavy scalar and the noHiggs amplitude is larger in absolute value than the
interference between the two Higgs. As a consequence the sum is non zero and agrees in sign with the
former of the two interferences but, as already mentioned, for parameters outside their allowed range.

\begin{table}[tbh]
%\vspace{0.15in}
\begin{center}
%\hspace*{-2mm}
\begin{tabular}{|c|c|c|c|c|c|c|c|c|}
\hline
& \multicolumn{4}{|c|}{200 GeV $< M_{4l} <$ 1 TeV} & 
\multicolumn{4}{|c|}{$\vert M_{4l} - M_H\vert \;<$ 25 GeV}  \\
\hline
           $M_H$ (GeV), $s_\alpha$
        & \hspace*{2mm} $\sigma$ \hspace*{2mm} 
        &  \hspace*{1mm} $\sigma_{0h} $ \hspace*{1mm}  
        & $\sigma_{0h}{\scriptstyle +}\sigma_ H$  
        & \hspace*{1mm} $\sigma_{SM}$ \hspace*{1mm} 
        & \hspace*{2mm} $\sigma$ \hspace*{2mm}  
        & \hspace*{1mm}  $\sigma_{0h} $ \hspace*{1mm}    
        &  $\sigma_{0h}{\scriptstyle +}\sigma_ H$   
        & \hspace*{1mm} $\sigma_{SM}$ \hspace*{1mm} \\  
\hline
400, 0.3, 4l & 98.1  & 87.4 &  98.2  & 86.9 & 18.3  & 8.1  &  18.3 & 8.1 \\
\hline
600, 0.2, 4l  & 89.4  & 87.0 & 89.5  & 86.9 & 5.2  & 2.9  &  5.3 &  2.9 \\
\hline
600, 0.2,  2l2$\nu$  &  5931 & 5870  & 5931  & 5874 & 248  &  193 &  253 & 192\\
\hline
\end{tabular}
\end{center}
\caption{
Cross sections in ab 
at the LHC with a center of mass energy of 13 TeV. $\sigma_{SM}$ corresponds to $s_\alpha=0$.
The SM cross section in $\vert M_{4l} - M_H\vert \;<$ 25 GeV can be considered as the SM background to
the heavy Higgs.}
\label{table:sigma}
\end{table}

The vector bosons in the heavy Higgs decay, for all the masses we have considered,
are predominantly longitudinally polarized.
In order to preserve unitarity the leading term of the contributions to jj$V_LV_L$ production from vector
boson interactions and from Higgs exchange must cancel each other exactly in the large energy limit,
where vector and Higgs masses can be neglected.
The near perfect suppression we observe between $A_0$ and $A_h$, which results in a small interference
of the heavy Higgs with the rest of the amplitude, suggests that the cancellation
between $A_0$ and $A_h$ sets in already for invariant masses of the vector pair of a few hundred GeV,
provided the mixing angle is not too large.

Similar analysis have been performed for the GGF case in \rf{Kauer:2015hia}, highlighting a partial
cancellation between the interferences of the Heavy Higgs with the light one and with the continuum.

In \tbn{table:sigma} we show the cross section in attobarns for two mass intervals:  
200 GeV $< M_{4l} <$ 1 TeV, which roughly coincides with the range employed so far by the experimental
collaborations to set limits on the presence and couplings of additional scalars,
and $\vert M_{4l} - M_H\vert \;<$ 25 GeV, as an indication of the possible effects on an
analysis in smaller mass bins which requires high luminosity.
The corresponding cross sections for the $\ordQCD$ 4l processes are 222 ab in the 200 GeV-1 TeV
interval, 18.4 ab in the 375 GeV-425 GeV range, 5.4 ab in the 575 GeV-625 GeV range.
For the  2l2$\nu$jj final state the $\ordQCD$ cross sections are 30 fb between 200 GeV and 1 TeV,
and 580 ab in the 575 GeV-625 GeV interval.

We notice that $\sigma_{0h} \approx \sigma_{SM}(s_\alpha=0)$ in both intervals. The only difference
between the two results is that in the first case the Higgs couplings are scaled by $c_\alpha$.
Therefore, the off shell predictions are hardly affected by this modification.

The incoherent sum $\sigma_{0h}{\scriptstyle +}\sigma_ H$ agrees with the exact result in all cases also
when integrated over.

Even in the smaller interval $\sigma_{0h}$ gives a substantial contribution to VV production and must be
taken into account when searching for a heavy Higgs.

\section{Conclusions}
We have studied Higgs sector interference effects in Vector Boson Scattering at the LHC,
both in the Standard Model and its one Higgs Singlet extension as a prototype of theories in which more
than one neutral, \CP even, scalars are present.
We have concentrated on $p p \rightarrow jj\, l^+l^-l^{\prime +}l^{\prime -}$
and $p p \rightarrow jj\, l^+\bar{\nu}_l l^{\prime -}\nu_{l^\prime}$  production.
We have shown that large interferences among the different Higgs exchange channels are present in the
SM and that a production times decay approach fails to reproduce the off shell Higgs contribution.
In the 1HSM, there are additional interferences between the two Higgs fields.
Different approximations have been tried and proved inaccurate.
We have found that the interference between the heavy Higgs diagrams and the rest of the amplitude, which is
the sum of
light Higgs exchange diagrams and of those diagrams in which no Higgs appear, is very small for values
of the mixing angle compatible with the experimental constraints and can be neglected.  

\section *{Acknowledgments}

Several stimulating discussions with Anna Kropivnitskaya and Pietro Govoni are
gratefully acknowledged.
This work has been supported by MIUR (Italy) under contract 2010YJ2NYW\_006,
by the Compagnia di San Paolo under contract  ORTO11TPXK and by the European
Union Initial Training Network HiggsTools (PITN-GA-2012-316704).\\

%\hfill*
%\eject
%\newpage

%%%%%%%%%%%%%%%%%%%%%%%%%%%%%%%%%%%%%%%%%%%%%%%%%%%%%%%%%%%%%%%%%%%%%%%%%

% To include bibliography do:
% 1- pdflatex FileName.tex
% 2- bibtex FileName
% 3- pdflatex FileName.tex
% 4- pdflatex FileName.tex
%\bibliographystyle{unsrt}
%\bibliographystyle{hunsrt}
%\bibliographystyle{h-elsevier.bst}

\bibliographystyle{JHEP}

\bibliography{MySinglet2}

\end{document}